\documentclass[fleqn,usenatbib]{mnras}

\usepackage{newtxtext,newtxmath}

\usepackage[T1]{fontenc}

\DeclareRobustCommand{\VAN}[3]{#2}
\let\VANthebibliography\thebibliography
\def\thebibliography{\DeclareRobustCommand{\VAN}[3]{##3}\VANthebibliography}

\usepackage{graphicx}	
\usepackage{amsmath}	
\usepackage{multirow}
\usepackage{xspace}

\newcommand{\swift}{{\em Swift}\xspace}
\newcommand{\fermi}{{\em Fermi}\xspace}
\newcommand{\decay}{{$\alpha_{X,>200s}$}\xspace}
\newcommand{\lum}{{$L_{\mathrm{X,200s}}$}\xspace}

\title[A correlation in GRB X-ray afterglows]{The intrinsic luminosity--decay correlation in subsamples of GRB X-ray afterglows}

\author[S. P. R. Shilling et al.]{S. P. R. Shilling$^{1,2,3}$\thanks{E-mail: s.shilling@lancaster.ac.uk},
S. R. Oates$^{1}$,
J. L. Racusin$^{2}$,
B. Cenko$^{2}$,
R. Gupta$^{2}$\thanks{NASA Postdoctoral Program Fellow},
P. Nuessle$^{2,4}$,
M. Smith$^{1}$,\newauthor
G. P. Lamb$^{5}$
and C. Turnbull$^{5}$
\\
$^{1}$ Department of Physics, Lancaster University, Lancaster, LA1 4YB, UK\\
$^{2}$ Astrophysics Science Division, NASA Goddard Space Flight Center, Mail Code 661, Greenbelt, MD 20771, USA\\
$^{3}$ Center for Research and Exploration in Space Science and Technology, NASA Goddard Space Flight Center, Greenbelt, MD 20771, USA\\
$^{4}$ Department of Physics, The George Washington University, 725 21st St. NW, Washington, DC 20052, USA\\
$^{5}$ Astrophysics Research Institute, Liverpool John Moores University, IC2 Liverpool Science Park, 146 Brownlow Hill, Liverpool, L3 5RF, UK\\
}

\date{Accepted XXX. Received YYY; in original form ZZZ}

\pubyear{2026}

\begin{document}
\label{firstpage}
\pagerange{\pageref{firstpage}--\pageref{lastpage}}
\maketitle

\begin{abstract}
The intrinsic luminosity--decay correlation in gamma-ray burst (GRB) afterglows, between the early-time luminosity and the average rate of decay past this time, has previously been observed in the radio, optical/UV, X-ray and GeV wavebands and quantitatively shows that more luminous afterglows tend to have higher average rates of decay. We have compiled an updated sample of 427 X-ray afterglows with measured redshifts, observed with \swift/XRT over 20 years. For each GRB, we measure the luminosity at 200 seconds in the rest frame, \lum, and the average rate of decay from this time, \decay. We find these parameters are correlated with a Spearman's rank coefficient ($R_\mathrm{sp}$) of $0.54\pm0.04$ at a significance of $\geq3\sigma$ and a linear regression slope of $0.18\pm0.02$. We separate our sample into subsamples, including 395 long GRBs (LGRBs) and 32 short GRBs (SGRBs) and find evidence of the \lum--\decay correlation at a significance of $\geq3\sigma$ in LGRBs but not in SGRBs, consistent with previous studies. In a subsample of 102 LGRBs with well-sampled light curves and late end times, we find that scatter in the correlation is significantly reduced and the strength increases to $R_\mathrm{sp}=0.80\pm0.04$ whilst the slope remains consistent with the full sample. We discuss our results and, briefly, their potential implications on constraining the cause of the correlation. Possible causes include geometric effects due to the angle between the observer and the jet-axis, or some mechanism that regulates the rate at which energy is released by the GRB central engine.
\end{abstract}



\section{Introduction}
\label{sec:introduction}
Gamma-ray bursts (GRBs) are short-lived astrophysical transient phenomena characterized by a brief episode (typically on the scale of seconds) of energetic gamma-ray emission known as the `prompt' emission accompanied by a longer-lasting phase (typically lasting up to a week but in some cases up to months or years) of multi-wavelength emission known as the `afterglow' \citep{Cavallo_78,pac86,goo86,she90,pac90,rees92,mes93,mes97,sar98}. 

For over three decades, GRBs have been been classified as either long-- or short--duration based on their measured $T_{90}$ duration: the time over which 90 per cent of their gamma-ray emission is detected, with the boundary at $T_{90} \simeq2$ s \citep{kou93}. Long GRBs (LGRBs) are typically spectrally soft and believed to be associated with collapsar progenitors and thus a supernova is expected to be observed coincident with the LGRB \citep{woo93,mac99}. Short GRBs (SGRBs) are spectrally hard, thought to be associated with compact mergers, and thus a kilonova, and possibly a gravitational-wave counterpart, are expected to be observed coincident with the SGRB \citep{pac86,eic89}. Strong evidence for the collapsar and merger progenitors were first presented with the coincident detections of LGRB 980425 with SN 1998bw \citep{gal98}, and SGRB 170817A with GW 170817A \citep[e.g.][]{gal98,abb17a,abb17b,cou17,eva17,gol17}. Thereby, it is implicitly assumed that long-- and short--duration GRBs are physically unique events corresponding to two different progenitors. 

However, as highlighted in \cite{Bromberg_2013}, classifying GRBs by their $T_{90}$ duration method is imperfect due to the systematic overlap between the durations of each class and the detector dependence of the duration itself, and can misclassify long and short GRBs. Examples of GRBs which were misclassified based on their $T_{90}$ duration include GRBs 200826A and 211211A. On the one hand, GRB 200826A was misidentified as a SGRB due to its short duration, even though its high energy, soft spectrum and collapsar origin are all consistent with typical LGRBs \citep{aham21,zha21_nat,ros22}. On the other hand, GRB 211211A was misidentified as a LGRB due to its long duration, even though its low energy, initially hard spectrum, merger origin, and coincident kilonova are all characteristic of typical SGRBs \citep{ras22,tro22,yan22}. 

There are also GRBs of each class (long and short) with properties that are sufficiently different enough to their respective GRB class as to warrant the need for a subclass. Some GRBs that are consistent with LGRBs by having a long duration and an associated supernova (indicating a collapsar origin) can also have a significantly lower isotropic energy release in the $\gamma$-ray band compared to other LGRBs. This invokes the need for the subclass of low-luminosity LGRBs \citep[e.g. GRB 060218][]{fer06}. Similarly, some GRBs can be largely consistent with SGRBs but can have an extended period of gamma-ray emission which is uncharacteristic of the typical SGRB population. These types of SGRBs are classified into a subclass known as SGRBs with extended emission \citep[e.g. GRB 181123B][]{dic21}. 

Correlations between GRB parameters, from both the prompt emission and the afterglow \citep[see][for a review]{dai17}, may be utilized as diagnostic tools, potentially providing both an additional means of classifying long and short GRBs, and a route to understanding GRBs. Correlations discovered in the prompt emission of GRBs include the "Amati" relation \citep{ama02} between the intrinsic spectral peak energy, $E_{p,i}$, and the isotropic-equivalent energy, $E_{iso,\gamma}$; and the "Yonetoku" relation \citep{yon04} between $E_{p,i}$ and the isotropic peak luminosity, $L_{p,i}$. A notable example in the afterglow is the "Dainotti" relation \citep{dai08,dai13,dai20,levine20} between the luminosity at the end time of the plateau, $L(T_a)$, and the end time of the plateau itself, $T_a$. 

This paper investigates the luminosity--decay correlation, previously observed across the electromagnetic spectrum \citep[specifically in the GeV; X-ray; optical/UV; and radio wavebands][respectively]{Hinds23,Racusin16,oates12,shilling25}. This correlation provides quantitative evidence that afterglows which are initially brighter tend to decay on average more quickly. In each previous study, the luminosity--decay correlation has only been observed in LGRBs. However, only one previous study \citep[in the X-ray band;][]{Racusin16} included a sample of SGRBs in their analysis and the sample size was limited, consisting of just 9 events, so it was not possible to confidently determine if the correlation existed in the SGRB population.

In this paper, we have compiled an updated sample of X-ray afterglows observed with \swift/XRT over 20 years, covering double the mission operation time compared to the previous X-ray study \citep{Racusin16}. With this updated sample, we re-examine the luminosity--decay correlation in X-ray afterglows and explore the correlation in different GRB subsamples. The sample selection, data analysis, and subsample selections are given in Section \ref{sec:analysis}; the results are given in Section \ref{sec:results}; a comparison of the results, the possible causes of the correlation, and applications of the correlation are discussed in Section \ref{sec:discussion}; and conclusions are given in Section \ref{sec:conclusions}. For this study we assume cosmological parameter values of $H_0 = 70$ km s$^{-1}$ Mpc$^{-1}$, $\Omega_{\Lambda} = 0.7$ and $\Omega_{m} = 0.3$, and the flux convention of $F(t,\nu) \propto t^{-\alpha} \nu^{-\beta}$, where $\alpha$ and $\beta$ are the temporal and spectral indices, respectively.

\section{Data analysis}
In this section we discuss the initial sample selection, the conversion of each light curve to the rest frame, the measurement of the correlation parameters (the luminosity at 200 s in the rest frame and the average rate of decay from this time), and how we classify GRBs into each subclass.

\label{sec:analysis}
\subsection{Initial sample selection}
\label{sec:initial sample}
We use the \texttt{swifttools} API\footnote{https://www.swift.ac.uk/API/} in \texttt{python} to retrieve the \swift/XRT count rate light curve, excluding upper limits, from the XRT repository\footnote{https://www.swift.ac.uk/xrt\_curves/} \citep{eva07,eva09} for every \swift/BAT triggered GRB between December 2004 and November 2024 \cite[as per the third BAT catalogue general summary][]{lie16}\footnote{https://swift.gsfc.nasa.gov/results/batgrbcat/index\_tables.html} with a \swift/XRT detection. Of these, we select the 552 that have a measured redshift in the literature to enable us to convert these light curves to the rest frame.

\subsection{Light curves}
\subsubsection{Converting to the rest frame}
\label{sec:rest frame}
For each individual GRB, we retrieve the photon index, $\Gamma=\beta+1$, and the single counts-to-unabsorbed-flux conversion factor, ECF, from the XRT repository. These are derived from the automated spectral fits to the time-averaged photon counting data over the entire \swift/XRT light curve, and do not account for spectral evolution. We convert each light curve from counts to unabsorbed flux ($\mathrm{erg\ cm^{-2}\ s^{-1}}$) in the \swift/XRT native band ($0.3-10$ keV) using the ECF. We then convert the light curves from flux in the $0.3-10$ keV band to flux density at 1 keV ($\mathrm{erg \ cm^{-2}\ s^{-1} \ Hz^{-1}}$) according to the following formula \citep[c.f.][]{geh08}:
\begin{equation}
\label{equation:density}
F_{[E_\nu]}=f_{[E_1,E_2]}\cdot4.13\times10^{-18} \cdot \left( \frac{E_{\nu}^{1-\Gamma}(2-\Gamma)}{E_2^{2-\Gamma} - E_1^{2-\Gamma}}  \right),
\end{equation}
where $F_{[E_\nu]}$ is the flux density ($\mathrm{erg \ cm^{-2}\ s^{-1} \ Hz^{-1}}$) at energy $E_\nu$ and $f_{[E_1,E_2]}$ is the flux ($\mathrm{erg\ cm^{-2}\ s^{-1}}$) in the $[E_1,E_2]$ energy bandpass. The native bandpass of \swift/XRT corresponds to $E_1=0.3$ keV and $E_2=10$ keV, and to calculate the flux density at 1 keV, we set $E_\nu=1$ keV. We then convert each light curve from the flux density at 1 keV to the \textit{k}-corrected \citep[see][]{blo01} intrinsic luminosity at 1 keV (erg s$^{-1}$ Hz$^{-1}$) in the GRB co-moving frame using:
\begin{equation}
\label{eqn:luminosity}
L=F_{[E_\nu]} \cdot4 \pi d_{L}^2 (1+z)^{\beta -1},
\end{equation}
where $L$ is the luminosity, $d_L$ is the luminosity distance, and $z$ is the redshift. We retrieve the redshift for each GRB from the literature as shown in Table \ref{tab:measurements}. Similarly, we convert each bin from the observed frame, ${t_\mathrm{obs}}$, to the rest frame, $t_{\mathrm{rest}}$, using:
\begin{equation}
\label{eqn:time}
t_{\mathrm{rest}}=\frac{t_\mathrm{obs}}{1+z}.
\end{equation}

\begin{center}
\begin{figure}
\includegraphics[width=\columnwidth]{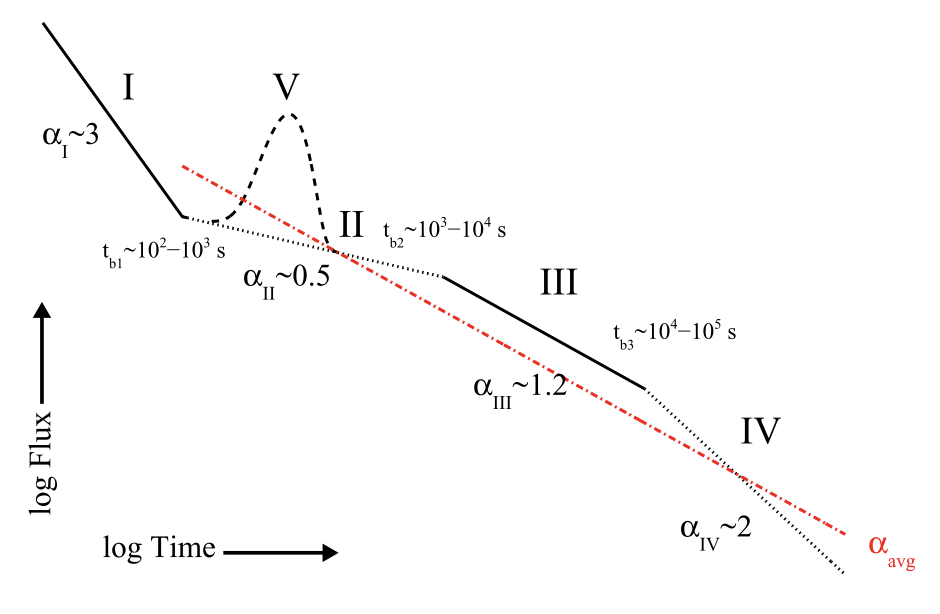}
    \caption{The `canonical' GRB X-ray afterglow light curve morphology \protect\citep{nousek,zhang06} in comparison to the average afterglow, $\alpha_\mathrm{avg}$, reproduced with permission from Figure 1 in \protect\cite{Racusin16}. Several common unique features can be identified: `0' denotes the prompt emission, `I' denotes the steep decay, `II' denotes the plateau, `III' denotes the forward shock decay, `IV' denotes the post-jet break decay, and `V' denotes a single flare episode. Steep decay segments and flares are related to the prompt emission and can dominate over the features attributed to the afterglow emission. }
    \label{fig:canonical}
\end{figure}
\end{center}

\subsubsection{Measuring the luminosity and average rate of decay}
\label{sec:measurements}
We aim to measure the luminosity at 200 s in the rest frame, \lum, and the average rate of decay from this time, \decay, following \cite{oates09,oates15} and \cite{Racusin16}. We wish to probe the pure afterglow phase; however, GRB X-ray light curves exhibit a complex structure, as illustrated in Figure \ref{fig:canonical}, consisting of up to five unique features including two -- the steep decay segment and highly variable flares -- that are believed to be related to the prompt emission \citep{nousek,zhang06}. These features can dominate over the afterglow emission. Reducing contamination of the afterglow by the prompt emission was shown to improve the luminosity--decay correlation by reducing scatter and increasing the strength of the Spearman's rank coefficient in \cite{Racusin16}. As such, we follow this method and attempt to reduce the contamination of the afterglow by the prompt emission by excluding these data from our measurements of \lum and \decay.

The time intervals of flare episodes (referred to as `flare intervals' from here onwards) are automatically identified by the \swift/XRT processing pipeline and are reported in the XRT catalogue\footnote{https://www.swift.ac.uk/xrt\_live\_cat/} along with the corresponding best-fit light curve model \citep{eva09}. For each light curve, we retrieve the best-fit model and the corresponding flare intervals from the XRT catalogue in addition to the morphological classification. Data within flare intervals are already excluded from these models \citep[see][]{eva07,eva09}; thus, the models characterize the remaining canonical features (i.e. features I-IV in Figure \ref{fig:canonical}), including any steep decay segments. We then identify steep decay segments in each light curve using the parameters from the best-fit models according to the following criteria: a temporal index of $\alpha > 2.5$ occurring at $t < 5000$ s, or a temporal index transition across a break of $\alpha_{n+1}-\alpha_n\leq -1$, where $n$ is the number of the power-law segment corresponding to each temporal index across each break. The first criterion is motivated by the typical canonical values of $\alpha\geq3$ (reduced to $\alpha\geq2.5$ to include some additional margin for error) reported in \cite{nousek,zhang06}. The latter criterion is motivated by the possible case of slope values of $\alpha < 2.5$ and thus not passing our first criterion but are steep relative to the rest of the light curve nonetheless and would not be identified using a flat slope threshold alone. For example; GRBs 160227A and 230116D, which have a transition in slope values across a break (from $\alpha_n$ to $\alpha_{n+1}$) of 1.5 to 0.2, and 1.8 to 0.62, respectively.

Having identified any steep decay segments and flare intervals, we proceed to select light curves and to measure the correlation parameters (\lum and \decay). Similarly to \cite{Racusin16}, we select light curves that were observed by \swift/XRT within 400 s of the \swift/BAT trigger in the rest frame to prevent extrapolation of fits from late times to 200 s in the rest frame. Additionally, we select light curves which have at least three bins at $t\mathrm{_{rest}}\geq200$ s to ensure a single power-law fit can be reasonably constrained. For any light curves with a steep decay segment or any number of flare intervals, we select those with three bins at $t\mathrm{_{rest}}\geq200$ s in the rest frame after the data within these features are excluded from the light curve, since they are excluded from our measurements of \lum and \decay. These selections reduce our sample size from 552 to 447 GRBs. 

To measure \lum, we evaluate the best-fitting light curve models at 200 s in the rest frame. As previously mentioned, flare intervals are already excluded from the light curves when these models were fitted. We therefore only need correct for steep decay contamination in the case of a steep decay segment ending after 200 s in the rest frame. There are 259 GRBs in our sample with a steep decay segment identified, of which there are 95 cases that require correction (lasting until after 200 s in the rest frame). We achieve these corrections by extrapolating the power-law segment from the best-fit model after the end of the steep decay segment (usually resembling the plateau), back to 200 s in the rest frame. We then measure \lum by evaluating the best-fit model at 200 s in the rest frame, using the extrapolated segment. 

\begin{center}
\begin{figure}
\includegraphics[width=\columnwidth]{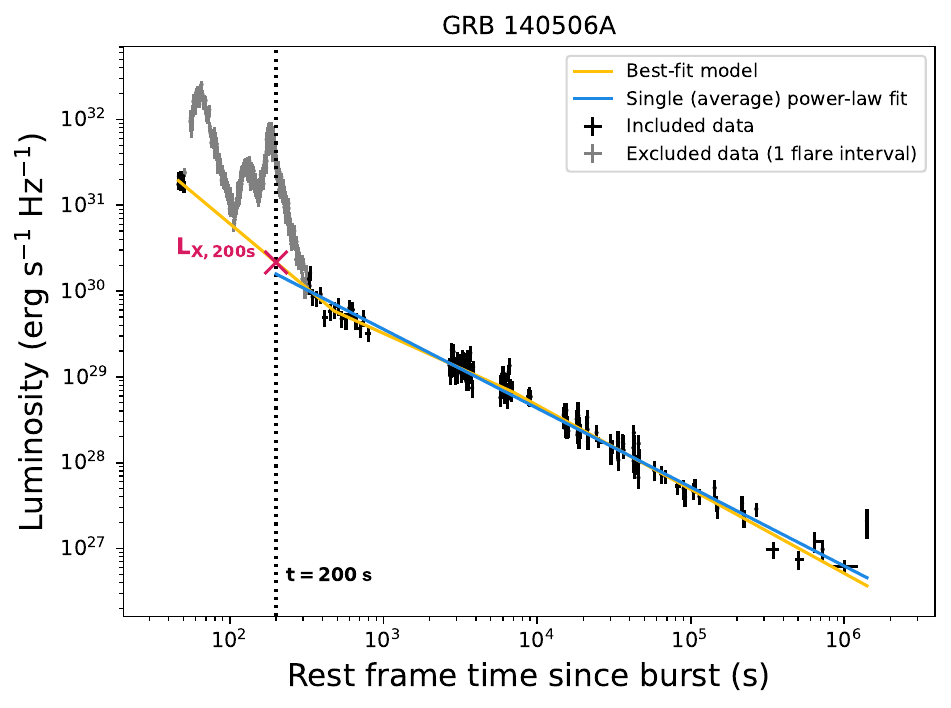}
    \caption{The rest-frame light curve of GRB 140506A. The best-fit model from the XRT catalogue, used to measure \lum, and the single power-law fit to the data from $t_\mathrm{rest}\geq200$ s, used to measure \decay, are overlaid. The light curve of GRB 140506A is contaminated by a flare interval. To correct for this contamination when measuring \lum and \decay, data within the flare interval are excluded from both fits.}
   \label{fig:light_curve_example_1}
\end{figure}
\end{center}

\begin{center}
\begin{figure}
\includegraphics[width=\columnwidth]{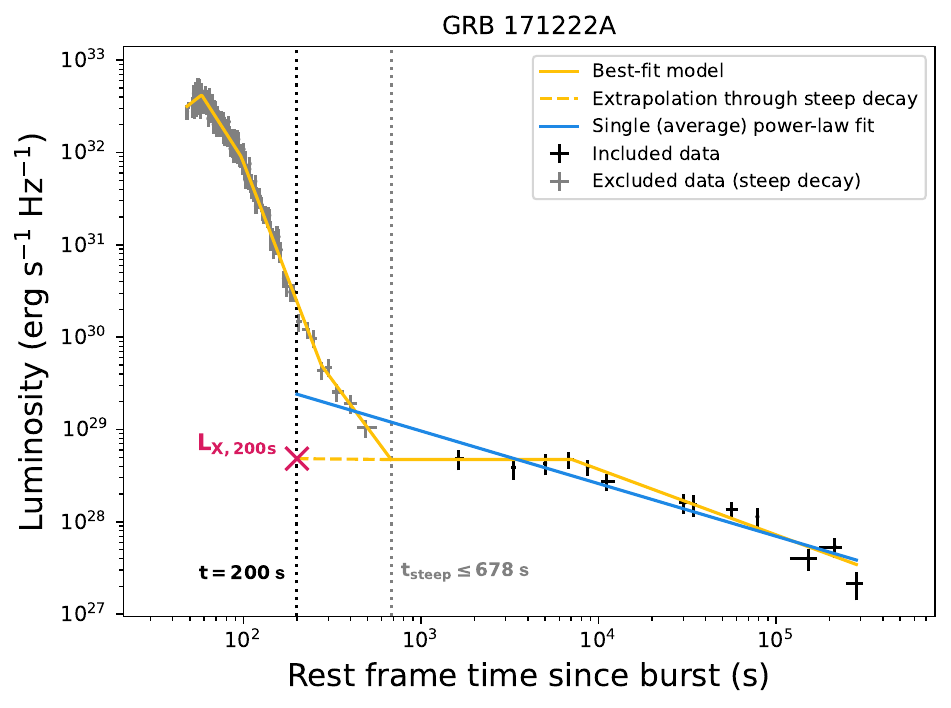}
    \caption{The rest-frame light curve of GRB 171222A. The best-fit model from the XRT catalogue, used to measure \lum, and the single power-law fit from $t_\mathrm{rest}\geq200$ s, used to measure \decay, are overlaid. The light curve of GRB 171222A is contamined by the tail end of the prompt emission at $t_\mathrm{rest}>200$ s via the steep decay segment. To correct for this contamination, we measure \decay by fitting the single power-law to the data from the end of the steep decay segment, from 678 s in this case, effectively excluding the data prior to this, and we measure \lum by extrapolating the power-law segment of the best-fit model after the end of the steep decay segment back to 200 s in the rest frame.}
   \label{fig:light_curve_example_2}
\end{figure}
\end{center}

To measure \decay, we fit a single power-law to each light curve from $t_\mathrm{rest}\geq200$ s using the \texttt{lmfit} module in \texttt{python}. When fitting, we correct for any prompt emission contamination by excluding data within any flare intervals reported in the XRT catalogue. In the 95 cases of a steep decay segment ending after 200 s in the rest frame, we instead fit a single power-law to the data from the end of the steep decay segment onwards. Figures \ref{fig:light_curve_example_1} and \ref{fig:light_curve_example_2} show the rest frame light curves of two GRBs which have contamination from a flare interval and a steep decay segment, with their fits overlaid, providing a visual example of how \lum and \decay are measured in each of these cases.

As highlighted in \cite{eva09}, the detection of flare intervals reported in the XRT catalogue are automated and may be misidentified. Following this, we inspect the flare intervals in each light curve to identify any possible misidentified cases as those which either: cover a large portion of the entire light curve, or occur at very late times ($t\mathrm{_{rest}}\geq10^4$ s) and do not appear highly variable in luminosity (thus not appearing to deviate from a power-law behaviour). We identify and exclude 8 such light curves, further reducing our sample size from 447 GRBs to 439. The light curves excluded here are shown, with their fits overlaid, in Appendix \ref{sec:appendix extra light curves}. Furthermore, we exclude any GRBs with a measured \lum uncertainty of $>2$ dex, or a measured \decay uncertainty of $>0.25$, finally reducing our sample size from 439 GRBs to 427. 

Throughout this paper, N denotes the number of GRBs in a given sample. The measurements of \lum and \decay for each GRB in our full sample, N=427, along with their redshifts and other parameters used in this analysis, are given in Table \ref{tab:measurements}.

\subsection{Prompt emission parameters}
\label{sec:prompt_params}
In order to classify GRBs, we require properties measured from the prompt emission, such as the $T_{90}$ duration and the isotropic-equivalent gamma-ray energy. We retrieve the measured $T_{90}$ duration for each GRB from the third BAT catalogue. We calculate the \textit{k}-corrected isotropic gamma-ray energy in the co-moving `bolometric' bandpass of $1-10,000$ keV, $E_{iso,\gamma}$, using the method in \cite{blo01}:
\begin{equation}
    E_{iso,\gamma}=\frac{4\pi D_L^2}{1+z} \cdot S_{[e_1,e_2]} \cdot k,
	\label{eqn:eiso}
\end{equation}
where $S_{[e_1,e_2]}$ is the observed fluence measured in the \swift/BAT native bandpass of $e_1=15$ keV and $e_2=150$ keV, and $k$ is the $k$-correction factor to the 1-10,000 keV band in the GRB co-moving frame. The calculation of the \textit{k}-correction factor is given in Appendix \ref{sec:appendix k-corr}, and we retrieve the measurements of $S_{[e_1,e_2]}$ from the third BAT catalogue using the time-averaged spectra produced from photons in the $T_{100}$ range: the time over which 100 per cent of the counts above the background are detected. The $E_{iso,\gamma}$ values, \textit{k}-corrected to the 1-10,000 keV band, calculated for each GRB are given in Table \ref{tab:measurements}.

\begin{table*}
\centering
\caption{The measured properties for each GRB in our sample. Col. (1): The GRB name. Col. (2): The GRB measured redshift. Col. (3): The reference for the measured redshift. Col. (4): Measurement of \decay. Col. (5): Measurement of \lum in log space. Col. (6): The \swift/BAT measured $E_{iso,\gamma}$ value (\textit{k}-corrected to the 1-10,000 keV band in the GRB co-moving frame) in log space. Col. (7): The \swift/BAT measured $T_{90}$. Col. (8): Flag for steep decay contamination (Y=yes, N=no). Col. (9): Number of flare intervals. Col. (10): Allocated subsample.}
\label{tab:measurements}
\begin{tabular}{lccccccccr}
\hline
GRB & Redshift & Redshift ref. & \decay & $\mathrm{log}_{10}$(\lum) & $\mathrm{log}_{10}(E_{iso,\gamma})$ & $T_{90}$ & Steep & No. of flares & Subsample \\
 &  &  &  & [$\mathrm{{erg\ s^{{-1}}\ Hz^{{-1}}}}$] & [$\mathrm{{erg}}$] & [s] &  &  &  \\
(1) & (2) & (3) & (4) & (5) & (6) & (7) & (8) & (9) & (10) \\

\hline
050126 & 1.29 & \protect{\cite{ber05}} & 0.98 $\pm$ 0.13 & 29.54 $\pm$ 0.06 & 52.64 & 48.00 & N & 0 & LGRB: reg. \\
050219A & 0.21 & \protect{\cite{ros14}} & 0.75 $\pm$ 0.05 & 27.65 $\pm$ 0.11 & 51.97 & 23.81 & Y & 0 & LGRB: reg. \\
050315 & 1.95 & \protect{\cite{ber05b}} & 0.66 $\pm$ 0.02 & 29.58 $\pm$ 0.30 & 53.09 & 95.40 & Y & 0 & LGRB: reg. \\
050319 & 3.24 & \protect{\cite{fyn09}} & 0.83 $\pm$ 0.03 & 30.56 $\pm$ 0.01 & 53.05 & 151.58 & N & 1 & LGRB: reg. \\
050401 & 2.90 & \protect{\cite{wat06}} & 1.08 $\pm$ 0.02 & 31.06 $\pm$ 0.00 & 54.04 & 32.09 & N & 0 & LGRB: reg. \\
\hline
\end{tabular}
\begin{minipage}[c]{2\columnwidth}
    {(The full table is available in its entirety in machine-readable format online.)} 
\end{minipage}
\end{table*}

\begin{center}
\begin{figure}
\includegraphics[width=\columnwidth]{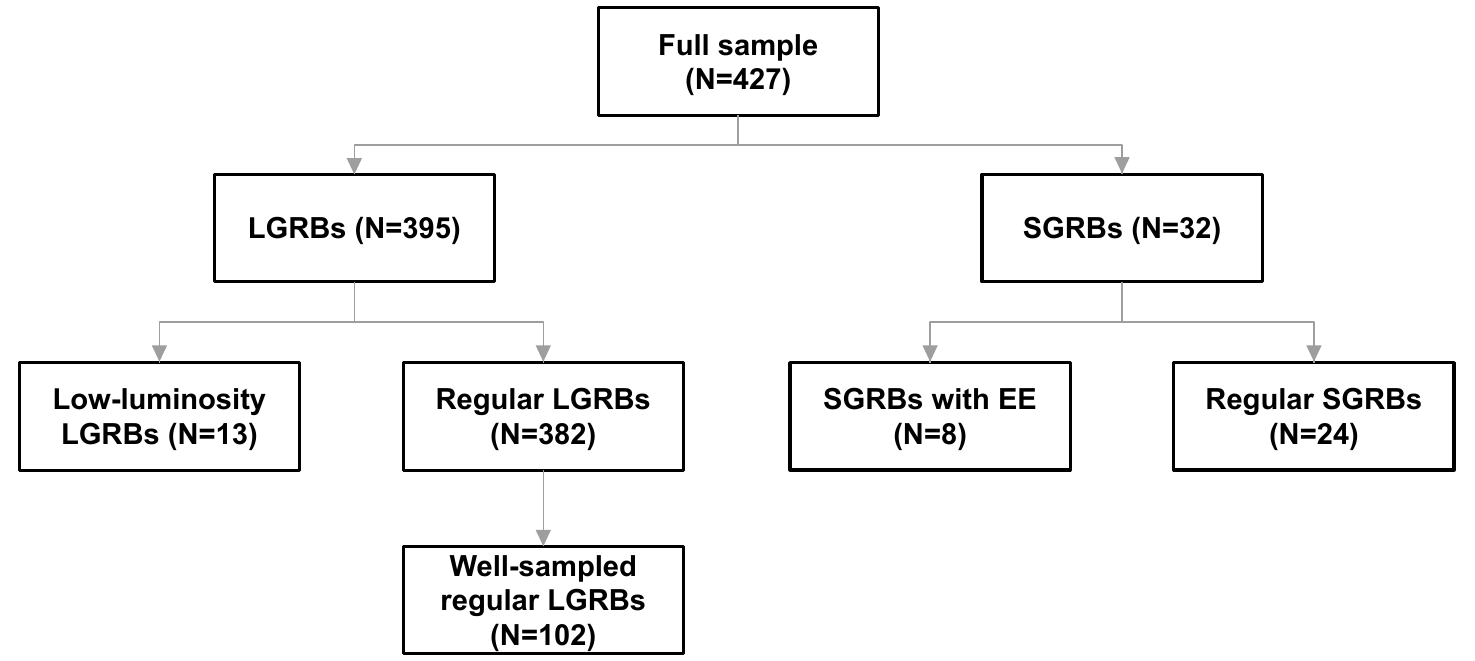}
    \caption{Flow chart summarising the number of GRBs in each subsample and their connection to previous parent samples. We use observed prompt emission properties to separate GRBs into long or short subsamples and then to further separate LGRBs into low-luminosity LGRBs or regular LGRBs, as well to further separate SGRBs into SGRBs with EE and regular SGRBs. We also classify regular LGRBs into a subsample of well-sampled regular LGRBs based on their light curve sampling. Each of these subsamples are analysed independently to test for the correlation.}
    \label{fig:flow_chart}
\end{figure}
\end{center}

\subsection{Subsample classification}
\label{sec:subsamples}
We first divide GRBs into two classes of long and short using their $T_{90}$ duration, with the boundary at $T_{90}= 2$ s \citep{kou93}. However, it is possible for some SGRBs to have a short pulse followed by an extended period of softer emission which can result in their $T_{90}\geq2$ s \citep[see GRBs 150424A, 111121A and 090916][]{lie16}. We therefore search the literature for definite cases of these and move any from our LGRB subsample into the SGRB subsample. 

We further separate our samples of LGRBs and SGRBs into subsamples. SGRBs are separated into those with and without extended emission (EE) using our list compiled from the literature. SGRBs without extended emission are classified as regular SGRBs. LGRBs are separated into those which are `low-luminosity' or not using a dividing boundary of $\mathrm{log}_{10}(E_{iso,\gamma})=51$ erg, $\sim2\sigma$ below the mean value of $52.94\pm0.90$ erg in our sample of LGRBs. We verify this threshold by confirming that GRBs 060218 and 100316D, two examples of known low-luminosity LGRBs \citep{fer06,sta11}, with $\mathrm{log}_{10}(E_{iso,\gamma})$ values of 49.23 and 50.07 erg respectively, are selected. Those that are not selected as low-luminosity LGRBs ($\mathrm{log}_{10}(E_{iso,\gamma})\geq51$ erg) are classified as regular LGRBs. We also select an additional subsample of regular LGRBs which have well-sampled light curves using the criterion of $\geq3$ data points in $\geq5$ rest frame time bins, each covering a single order of magnitude in log space, for instance $10^1-10^2$ s and $10^2-10^3$ s. These rest frame time bins in log space will be hereafter referred to as a `dex bins'. Further details on the selection of well-sampled light curves are given in Section \ref{sec:well-sampled}.

Figure \ref{fig:flow_chart} summarizes our subsample sizes and the allocated subsample for each GRB is given in Table \ref{tab:measurements}. A comparison of the rest frame light curve distributions between each subsample and the full sample is shown in Appendix \ref{sec:appendix extra light curves 2}. We examine the possible biases and choice of selection criteria for each subsample based on prompt emission properties in Section \ref{sec:instrumental_biases}.

\subsection{Correlation analysis}
For each subsample, we perform a Spearman's rank test, using the \texttt{scipy.stats} module in \texttt{python} \citep{scipy20}, between our measured values of the luminosity at 200 s in the rest frame and the average rate of decay from this time onwards. The Spearman's rank test calculates a Spearman's rank coefficient and a p-value, which respectively characterize the strength of the correlation and the likelihood that the correlation observed is due to chance. We additionally perform an error-weighted linear regression analysis of \lum and \decay using the \texttt{scipy.odr} module in \texttt{python}, which characterizes the relationship between these variables. 

We calculate the uncertainties on the Spearman's rank coefficient and the linear regression parameters via Monte Carlo Bootstrap calculations \citep{iso90, fei92}. For $10^{5}$ trials, we randomly resample, with replacement, our measurements of the luminosity at 200 s and the average rate of decay from 200 s onwards in pairs, to produce a simulated sample of the same size as our real sample. In each trial, we calculate the Spearman's rank coefficient and the linear regression parameters for the simulated sample, and record the results. After $10^{5}$ trials, the $\pm1\sigma$ lower and upper errors are calculated for each result by taking the difference between the mean of the simulated results and the 15.9th (lower) and the 84.1st (upper) percentile values of the simulated results, respectively.

\section{Results}
\label{sec:results}
We first test for the correlation in the full sample (long and short) and then in the individual samples of long and short GRBs. Next, we further separate long and short GRBs into their respective subclasses and test each individually for the correlation. We then examine the possible instrumental biases in each sample due to selecting subsamples based on prompt emission properties, since prompt emission properties are detector dependent. Finally, we present the final correlation and examine this for possible biases.

\begin{center}
\begin{figure}
\includegraphics[width=\columnwidth]{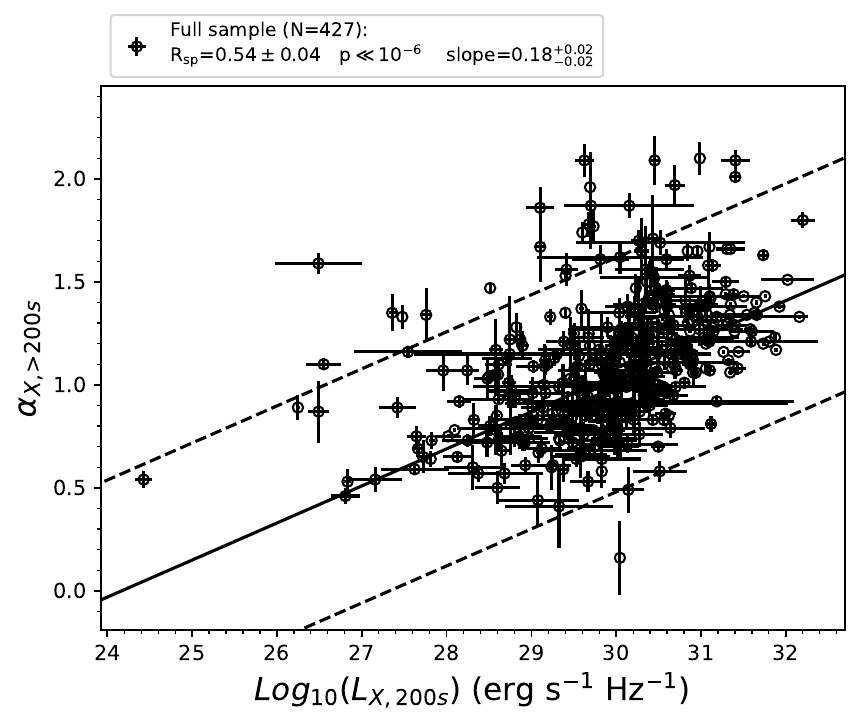}
    \caption{Plot of \decay vs \lum and the linear regression for our full sample. We observe the correlation in our full sample with $R_\mathrm{sp}=0.54 \pm 0.04$ at a confidence interval of $\geq3\sigma$. The solid line overlaid is the linear regression and the dashed lines are the $\pm2\sigma$ RMS (root-mean-square) deviation, which is also the case for the other correlation figures throughout this paper.}
   \label{fig:full_sample}
\end{figure}
\end{center}

\subsection{Subsamples}

\subsubsection{Long and short GRBs}
Before separating GRBs into long and short samples, we test the full sample for the correlation. Figure \ref{fig:full_sample} shows the measurements of \decay against \lum in the full sample (N=427). We calculate a Spearman's rank coefficient of $R_\mathrm{sp}=0.54 \pm 0.04$, indicating that \lum and \decay are correlated such that more luminous X-ray afterglows tend to decay on average more quickly than less luminous ones. We calculate an associated p-value of p $\ll10^{-6}$ suggesting that the correlation is observed at a confidence level of $\geq3\sigma$ and therefore is not due to statistical fluctuation. We calculate a linear regression slope of $0.18^{+0.02}_{-0.02}$.

\begin{center}
\begin{figure}
\includegraphics[width=\columnwidth]{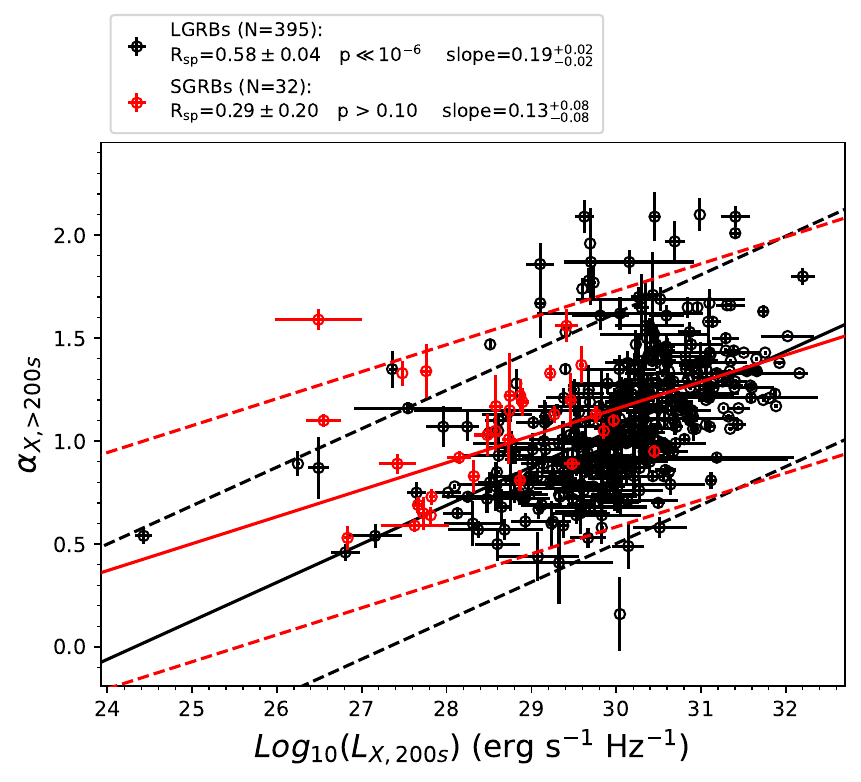}
    \caption{Plot of \decay vs \lum for LGRBs and SGRBs with the corresponding linear regression overlaid. A statistically significant correlation is observed in LGRBs but not in SGRBs.}
   \label{fig:long_and_short}
\end{figure}
\end{center}

We separate GRBs in the full sample into long and short classes. Figure \ref{fig:long_and_short} shows \decay against \lum for LGRBs (N=395) and SGRBs (N=32). We calculate a Spearman's rank of $R_\mathrm{sp}=0.58 \pm 0.04$, a p-value of p $\ll10^{-6}$, and a linear regression slope of $0.19^{+0.02}_{-0.02}$ for LGRBs and $R_\mathrm{sp}=0.27 \pm 0.20$, p $>0.10$ ($\mathrm{p}=0.72$), and a linear regression slope of $0.13^{+0.08}_{-0.08}$ for SGRBs. These results indicate that the correlation is observed in LGRBs, at a significance of $\geq3\sigma$, but there is no evidence for a correlation at this significance level in the SGRB sample.

\subsubsection{Low-luminosity LGRBs versus regular LGRBs}
\begin{center}
\begin{figure}
\includegraphics[width=\columnwidth]{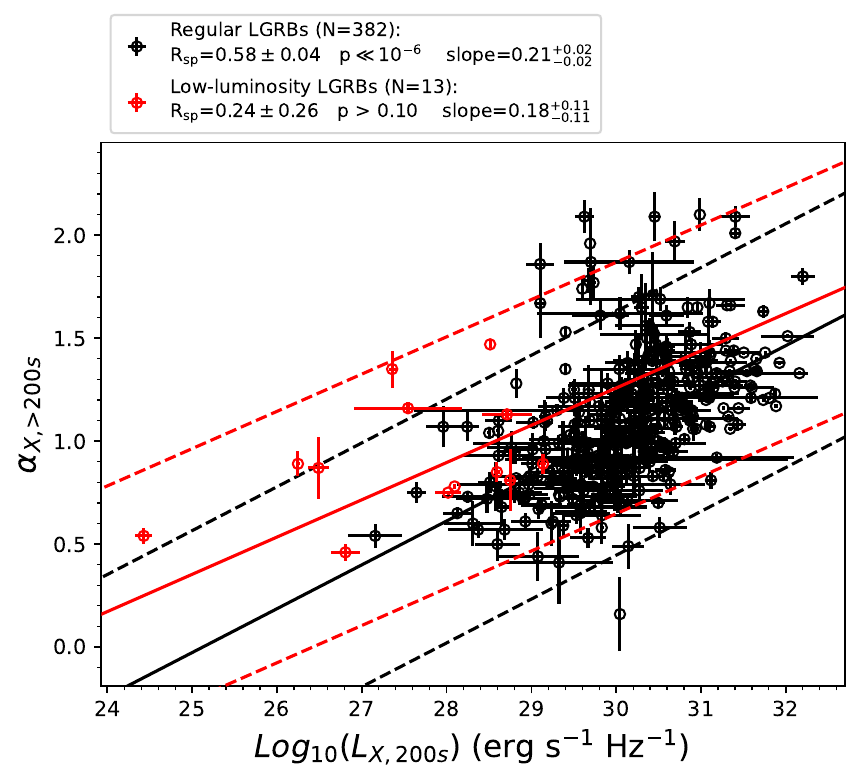}
    \caption{Plot of \decay vs \lum for the subsamples of low-luminosity LGRBs and regular LGRBs with the corresponding linear regression overlaid. A statistically significant correlation is observed in the subsample of regular LGRBs but not in low-luminosity LGRBs.}
   \label{fig:long_low_lum}
\end{figure}
\end{center}

We separate LGRBs into low-luminosity and regular LGRBs and find that the correlation is observed at a significance of $\geq3\sigma$ in regular LGRBs (N=382), with $R_\mathrm{sp}=0.58 \pm 0.04$ and p $\ll10^{-6}$, however we do not see evidence for a correlation in low-luminosity LGRBs (N=13), with $R_\mathrm{sp}=0.24 \pm 0.26$ and p $>0.10$ ($\mathrm{p}=0.43$). We note there are only 13 objects in the low-luminosity LGRB subsample, so it is unreliable to determine if the correlation exists in this subsample or not. The strength of the correlation in regular LGRBs is consistent with that of the parent LGRB sample, which is expected given there are only 13 fewer (low-luminosity) LGRBs. Figure \ref{fig:long_low_lum} shows a comparison of the correlation in low-luminosity LGRBs and regular LGRBs. 

\subsubsection{Well-sampled regular LGRBs}
\label{sec:well-sampled}
To identify a possible cause of scatter in the \lum--\decay correlation, we inspect light curves and their corresponding degree of scatter in the correlation. We note that light curves with less time coverage and sampling tend to result in more scatter in the correlation. Therefore, we establish a metric which represents the sampling and the time coverage of the light curve to investigate how the quality of the data affects the correlation. 

As mentioned in Section \ref{sec:subsamples}, we use the term `dex bin' to refer to a single order of magnitude of time in log space, for instance $10^1-10^2$ s and $10^2-10^3$ s. For each light curve, we count the number of rest frame time dex bins that contain $\geq3$ data points (to ensure a power-law can be reasonably weighted across each dex bin). Light curves with good sampling and a long duration will have a larger number of dex bins containing $\geq3$ data points. We include data within steep decay segments and flare intervals when counting the number of dex bins containing $\geq3$ data points, otherwise GRBs that have data in flare intervals or steep decay segments that are well-sampled will have a smaller number of dex bins to count over due to the exclusion of data. We select light curves that meet the criteria ($\geq3$ data points) in $\geq5$ dex bins\footnote{In other words, we require light curves to contain $\geq3$ data points in at least five of the following rest frame time slices: $10^1-10^2$ s, $10^2-10^3$ s, $10^3-10^4$ s, $10^4-10^5$ s, $10^5-10^6$ s, $10^6-10^7$ s, and $10^7-10^8$ s. We do not require the number dex bins containing $\geq3$ data points to be consecutive.}. We choose $\geq5$ dex bins because this optimizes the requirement for $\geq3$ data points up to a late time but not too late as to reduce the resulting sample size too greatly, as the requirement for more data at later times results in more GRBs being excluded. In Appendix \ref{sec: appendix well-sampled options} we discuss the selection criterion further and investigate how alternative choices for the number of dex bins, as well as how the inclusion of data in flare intervals and steep decay segments, affect the correlation.

\begin{center}
\begin{figure}
\includegraphics[width=\columnwidth]{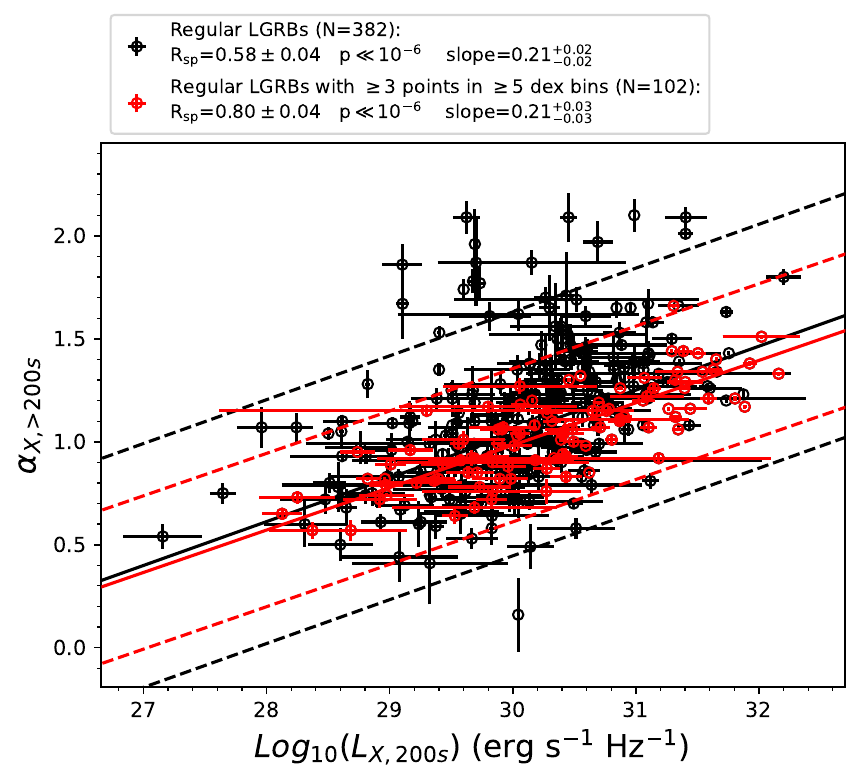}
    \caption{Plot of \decay vs \lum for the regular LGRB and well-sampled regular LGRB subsamples with the corresponding linear regression overlaid. A statistically significant correlation is observed in both subsamples, however the well-sampled regular LGRBs have significantly less scatter and a stronger Spearman's rank coefficient compared to the regular LGRBs.}
   \label{fig:lgrbs_by_data_quality}
\end{figure}
\end{center}

We proceed with selecting regular LGRBs with light curves with $\geq3$ data points in $\geq5$ dex bins resulting in a subsample of N=102 well-sampled regular LGRBs. We test for a correlation in this subsample and calculate $R_\mathrm{sp}=0.80 \pm 0.04$, p $\ll10^{-6}$, and a linear regression slope of $0.21^{+0.03}_{-0.03}$ as shown in Figure \ref{fig:lgrbs_by_data_quality}. Selecting well-sampled regular LGRBs according to our metric causes an increase in the correlation strength, as well as a significant reduction in the level of scatter, compared to the parent subsample of regular LGRBs. We also note that whilst the strength of the correlation increases, the slope of the linear regression is unchanged (to 2 decimal places) compared to the parent regular LGRBs. 

Scatter may arise in the correlation from a poorly sampled light curve with an early end time or large gaps in data which could cause the measurement of \decay to be misconstrued from the truth (as measured from a more informative light curve). A light curve could be poorly sampled due to changes in target observing priority or observing constraints including; anti-sun operation mode, the GRB becoming eclipsed by the Earth, or moon constraints. In turn, these can result in individual epochs of the afterglow's decay, such as the standard decay, having less coverage than a different phase, such as the plateau. Consequently, power-law fits will be more heavily weighted by an individual feature in the light curve's decay phase and are not accurately averaged across a reasonable range of light curve temporal behaviours. \cite{Racusin16} showed that the correlation is not driven by individual light curve segments. Therefore, it is important to span a large portion of the light curve to ensure a representative measurement of the average rate of decay is made across a range of temporal behaviours. It may be useful in a future paper to investigate how the sampling of a light curve affects the measurement of \decay.

\subsubsection{SGRBs with extended emission versus regular SGRBs}
\label{sec:outlier}
\begin{center}
\begin{figure}
\includegraphics[width=\columnwidth]{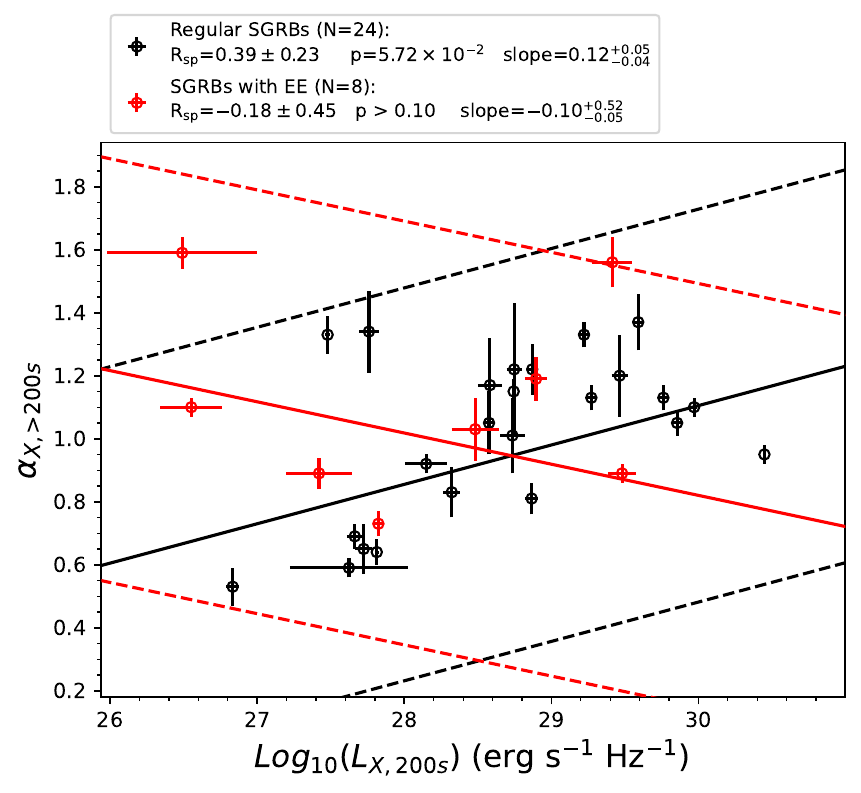}
    \caption{Plot of \decay vs \lum for the subsamples of SGRBs with extended emission and regular SGRBs, with the corresponding linear regression overlaid. No correlation is observed at a significance of $\geq3\sigma$ in either subsample.}
   \label{fig:short_subsamples}
\end{figure}
\end{center}

Figure \ref{fig:short_subsamples} shows a plot of \decay against \lum and the correlation results for our subsamples of SGRBs with EE (N=8) and regular SGRBs (N=24). For SGRBs with EE, we calculate $R_\mathrm{sp}=-0.18 \pm 0.45$ and a p-value of p $>0.10$ ($\mathrm{p}=0.67$). For regular SGRBs, we calculate $R_\mathrm{sp}=0.39 \pm 0.23$ and a p-value of $p=5.72\times10^{-2}$. These results suggest that, despite improved statistics to the subsample of regular SGRBs after separating them from SGRBs with EE, no statistically significant correlation is observed in either SGRB subsample. We note that these results may not give robust conclusions due to the sample size, and that the correlation analysis of a future larger sample will be useful in this regard.

\subsection{Testing for instrumental biases and selection effects}
\label{sec:instrumental_biases}
In Section \ref{sec:subsamples}, we separated GRBs into subsamples based on their observed properties for their individual correlation analyses. However, some properties are detector-dependent and therefore may be unreliable or subject to instrumental biases. There may also be selection effects associated with some classification methods. We therefore examine possible instrumental biases and selection effects and determine how these affect our subsample sizes and correlation results.

\subsubsection{The $T_{90}$ duration in selecting LGRBs and SGRBs}
We use the $T_{90}$ duration measured by the \swift/BAT to separate GRBs into short and long subsamples, with the boundary at $T_{90}=2$ s. However, the $T_{90}$ duration is a detector dependent property, since its value depends on the hardness of the detector energy bandpass. As highlighted in \cite{Bromberg_2013}, the $T_{90}$ distribution shows a dividing line (between long and short) at $T_{90}\simeq0.8$ s when measured by \swift/BAT (which covers $15-150$ keV) in comparison to the classical dividing line at $T_{90}\simeq2$ s when measured by BATSE \citep[which covers $30-1900$ keV][]{geh93}. Furthermore, it is often implicitly assumed that the different $T_{90}$ distributions correspond to physically different types of events, however, observations have shown that long GRBs can have a non-Collapsar origin and, conversely, short GRBs can have a Collapsar origin (see GRBs 200826A and 211211A; described in Section \ref{sec:introduction}). In turn, this means that classifying GRBs as long or short by using a fixed boundary on the $T_{90}$ duration can lead to the contamination of LGRBs by non-Collapsars and vice versa for SGRBs.

Due to the possible subsample contamination arising from the detector dependency of the $T_{90}$ duration, and the overlap between the durations of each population, we examine how classifying GRBs using the boundary of $T_{90}=2$ s affects our subsample sizes and correlation results. We re-select long and short GRBs using the shorter boundary of $T_{90}=0.8$ s and repeat the analysis. This results in 14 GRBs\footnote{GRBs 051221A, 070809, 090426, 100724A, 120804A, 121226A, 131004A, 140129B, 140930B, 150831A, 151229A, 170728A, 210323A, 211023B} being classified as long which were previously short. The $R_\mathrm{sp}$ values for the new ($T_{90}\gtrless0.8$ s) LGRB and SGRB samples are $0.56\pm0.04$ and $0.13\pm0.28$ respectively. These are both consistent at $\leq1\sigma$ with the values for their corresponding $T_{90}=2$ s boundary samples. Additionally, the changes in p-values in each respective sample using the shorter boundary are negligible. We therefore note that our main conclusions do not change if using a short-long boundary of $T_{90}=0.8$ s instead of $T_{90}=2$ s. 

\begin{center}
\begin{figure}
\includegraphics[width=\columnwidth]{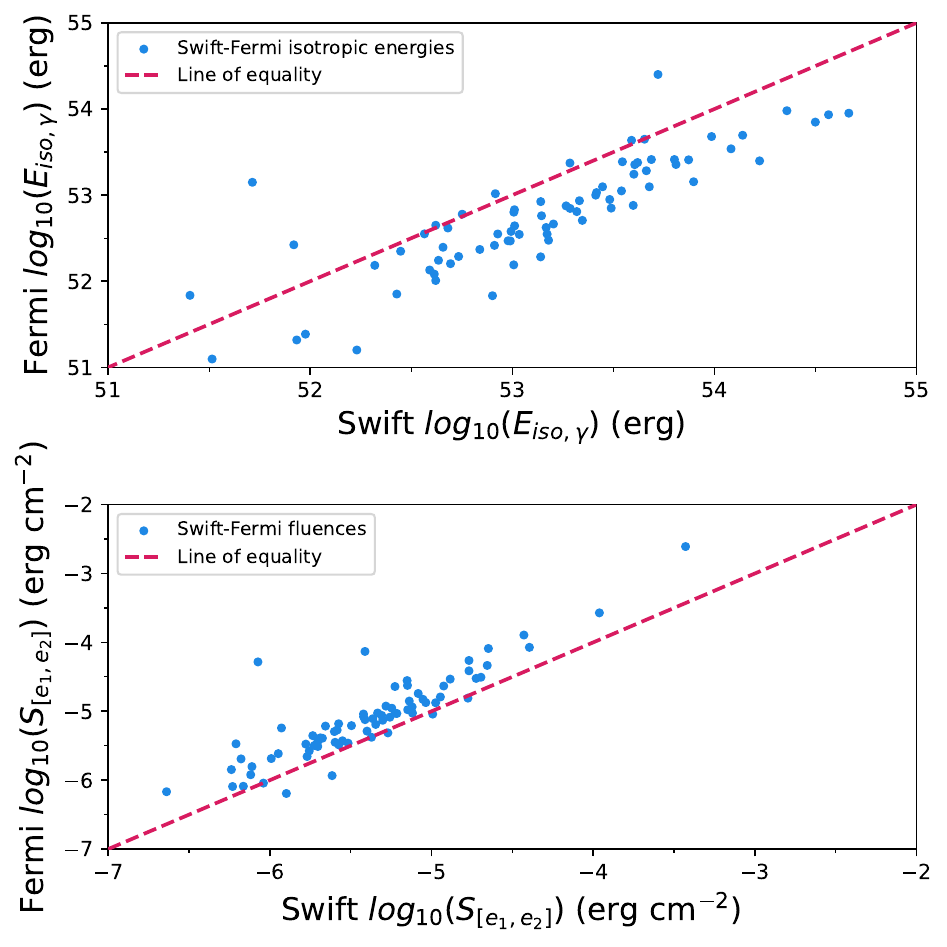}
    \caption{Comparison of the isotropic-equivalent energy k-corrected to the $1-10,000$ keV energy range (top) and the native band fluence (bottom) of \fermi/GBM and \swift/BAT for LGRBs in our sample that have been observed by both instruments. The native energy bandpasses are $e_1=8$ keV and $e_2=40$ MeV for \fermi/GBM (though the fluences from this instrument are measured between $10-1000$ keV) and $e_1=15$ keV and $e_2=150$ keV for \swift/BAT. \swift/BAT appears to systematically underestimate the fluence whilst overestimating the isotropic energy compared to \fermi/GBM. This may lead to the contamination of our regular LGRB sample with some low-luminosity LGRBs, but is unlikely to significantly affect the correlation results due to the large sample size of regular LGRBs.}
   \label{fig:fermi_eiso_comparison}
\end{figure}
\end{center}

\subsubsection{The isotropic gamma-ray energy in selecting regular and low-luminosity LGRBs}
We use the isotropic gamma-ray energy measured from the \swift/BAT spectra to classify LGRBs as low-luminosity or regular LGRBs. However, the relatively narrow energy range of \swift/BAT ($15-150$ keV) compared to other instruments, from different gamma-ray observatories, restricts its ability to constrain the peak energy, $E_{p}$. If $E_{p}$ is unknown, then the functional form of the spectrum is unknown outside the instrument bandpass energy range, and therefore the extrapolation of the spectral fit to a `bolometric' range of $1-10,000$ keV when calculating $E_{iso,\gamma}$ may be unreliable.

We examine if the $E_{iso,\gamma}$ values measured with \swift/BAT in its narrower bandpass of $15-150$ keV (\textit{k}-corrected to the `bolometric' $1-10,000$ keV range as detailed in Section \ref{sec:prompt_params}) are reliable by comparing them to those measured by \fermi/GBM, which has a larger bandpass of $8$ keV to $40$ MeV \citep{Mee09}. We retrieve the $E_\mathrm{iso,\gamma}$ values measured by \fermi/GBM for 121 LGRBs reported in \cite{po21}, of which there are 76 in common with our sample. These values have been \textit{k}-corrected \citep{blo01} to the `bolometric' $1-10,000$ keV range using the Band spectral model \citep{band93}. We also retrieve the \fermi/GBM fluences measured from the spectral fit between $10-1000$ keV, without requiring extrapolation, for these LGRBs from the Fermi GBM Burst Catalogue\footnote{https://heasarc.gsfc.nasa.gov/W3Browse/fermi/fermigbrst.html} for comparison with those measured by \swift/BAT in its narrower bandpass of $15-150$ keV. Figure \ref{fig:fermi_eiso_comparison} shows a comparison of these observed properties measured by each instrument for these 76 LGRBs.

As indicated by the lines of equality in Figure \ref{fig:fermi_eiso_comparison}, \swift/BAT appears to systematically underestimate the fluence whilst overestimating the isotropic-energy compared to \fermi/GBM. The underestimated fluence and an overestimated isotropic energy by \swift/BAT compared to \fermi/GBM are both likely due to the narrower native bandpass of \swift/BAT which captures less flux from higher energy photons, thus underestimating the fluence, and requires a larger extrapolation of the spectral model to $10-10,000$ keV without a measurement of $E_p$, thus overestimating the isotropic-equivalent energy. We examine how the systematic offset in $E_{iso,\gamma}$ values measured by \swift/BAT compared to by \fermi/GBM affects our selection of low-luminosity LGRBs.

The minimum \fermi/GBM measured $\mathrm{log}_{10}(E_{iso,\gamma})$ for the LGRBs in common with our sample (N=76) is $51.10$ erg, and thus these LGRBs are unlikely to contain any low-luminosity LGRBs. Instead, we re-select low-luminosity LGRBs from our full LGRB sample (N=395) using different thresholds on the \swift/BAT measured $\mathrm{log}_{10}(E_{iso,\gamma})$ value. We remind the reader that, in Section \ref{sec:subsamples}, when selecting low-luminosity LGRBs we used a threshold of $\leq51$ erg, which is $\sim2\sigma$ below the mean value for LGRBs. Here, we use $\leq50.5$ and $\leq51.5$ erg, corresponding to $\sim 2.5\sigma$ and $\sim1.5\sigma$ below the mean, which results in N=7 and N=23 low-luminosity LGRBs respectively. We analyse each new sample for the luminosity--decay correlation. For the $\mathrm{log}_{10}(E_{iso,\gamma})\leq50.5$ erg low-luminosity LGRBs (N=7), we calculate $R_\mathrm{sp} = 0.71\pm0.24$ and p-value of $7.13\times10^{-2}$ but note that this could be due to a small sample size of only 7 events. For the $\mathrm{log}_{10}(E_{iso,\gamma})\leq51.5$ erg low-luminosity LGRBs (N=23), we calculate a Spearman’s rank coefficient of $R_\mathrm{sp} = 0.40\pm0.19$, as well as a p-value of $5.86\times10^{-2}$. However, we note that the stronger Spearman's rank coefficient and the lower p-value (in comparison to our main sample of low-luminosity LGRBs shown in Figure \ref{fig:long_low_lum}) may be due to increased contamination of low-luminosity LGRBs by regular LGRBs, where the correlation is known to exist. 

A systematic overestimation of $\mathrm{log}_{10}(E_{iso,\gamma})$ by \swift/BAT is likely to result in the contamination of our sample of regular LGRBs by some low-luminosity LGRBs. Though, this effect is somewhat accounted for by using a threshold relative to the biased distribution, rather than a flat threshold determined by other unbiased instruments. Additionally, due to the large sample size of regular LGRBs (N=382) it is unlikely that some contamination will significantly affect the correlation results. The chance of contamination could be reduced significantly if measurements from an instrument with a wider bandpass (capable of measuring $E_p$) are used. 

\begin{center}
\begin{figure}
\includegraphics[width=\columnwidth]{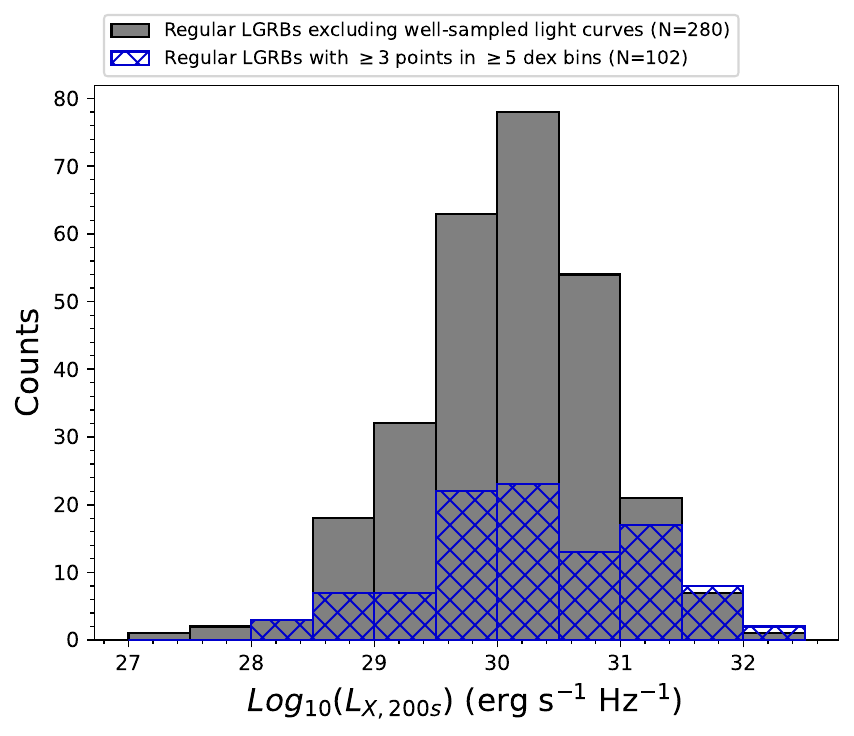}
    \caption{Distributions of the measured \lum values for our subsample of regular LGRBs divided into two groups: with and without well-sampled light curves. There is no visible offset between the distributions and we calculate a KS statistic of 0.17 with a p-value of $2.50\times10^{-2}$ verifying there is no bias at a confidence interval of $\geq3\sigma$. This suggests that the reduced scatter of the correlation in well-sampled regular LGRBs is not related to a possible bias in rest frame brightness.}
   \label{fig:lum_dist}
\end{figure}
\end{center}

\subsubsection{Testing for intrinsic biases in well-sampled regular LGRBs}
\label{sec:well_samp_biases_main}
The sampling of a light curve depends on many factors including the cadence, observing constraints, and the science plan, as well as the methods used to bin the data. The sampling also depends on the GRB brightness in the observed frame, because the binning depends on the number of counts per bin, which in turn depends on both the redshift and the intrinsic brightness. Given the multitude of possible factors related to a GRB having a well-sampled light curve or not, we test this subsample for biases to understand if they are physically different from regular LGRBs without well-sampled light curves. 

We separate regular LGRBs into two groups: with and without well-sampled light curves (meaning those containing $\geq3$ data points in $\geq5$ dex bins or not, as described in Section \ref{sec:well-sampled}). Figure \ref{fig:lum_dist} shows a comparison of the $\log_{10}$(\lum) distributions in each group. There is no visible offset between the distributions and we calculate a two-sample Kolmogorov-Smirnov (KS) statistic of 0.17 with a p-value of $2.50\times10^{-2}$, suggesting that there is no bias in \lum at a confidence interval of $\geq3\sigma$. The consistent rest frame X-ray brightnesses between the two groups indicates that scatter in the correlation is inherently due to light curves that are poorly sampled instead of a systematic difference in rest frame brightness.

\begin{center}
\begin{figure}
\includegraphics[width=\columnwidth]{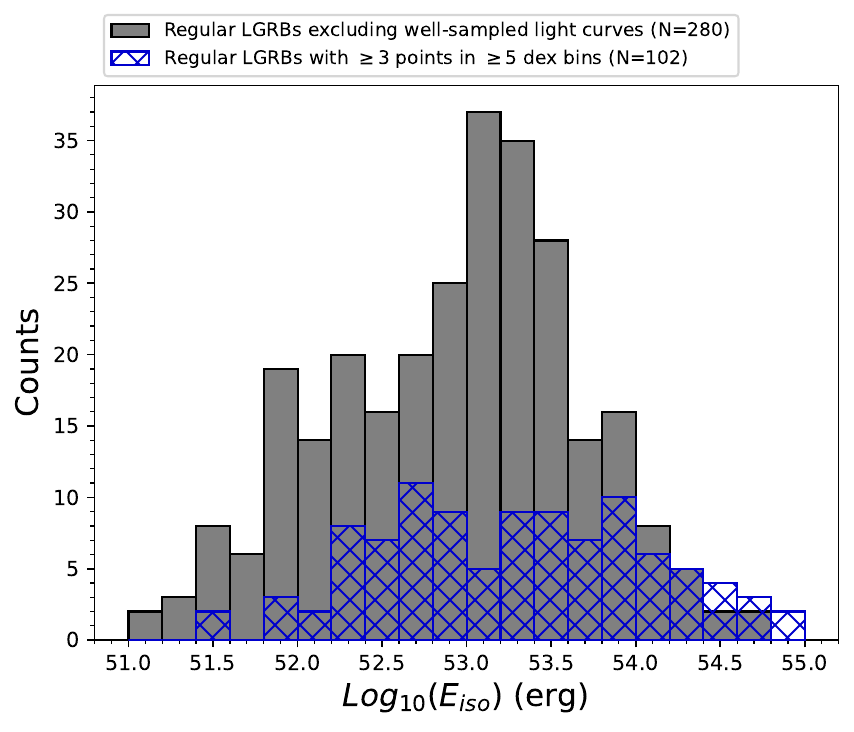}
    \caption{Distributions of the isotropic $\gamma$--ray energy for our subsample of regular LGRBs divided into two groups: with and without well-sampled light curves. The regular LGRBs with well-sampled light curves are energetically biased towards higher energies compared to regular LGRBs without well-sampled light curves. However, the KS statistic of 0.21 indicates the difference is small. Additionally, the ranges of each distribution are similar. Therefore, it is unlikely that the increased correlation strength, and reduction of scatter, observed in well-sampled regular LGRBs is due to an energetic bias.}
   \label{fig:eiso_dist}
\end{figure}
\end{center}

We examine the distributions of $\mathrm{E_{iso,\gamma}}$ in each group for a bias, as shown in Figure \ref{fig:eiso_dist}. There is a small degree of visible offset between the distributions and we calculate a KS statistic 0.21 and a p-value of $1.87\times10^{-3}$, indicating that, at a confidence interval of $\geq3\sigma$, well-sampled regular LGRBs are biased towards higher energies. However, the KS statistic is low and the ranges of each distribution are similar with a large degree of overlap, indicating the bias is subtle. Therefore, it is unlikely that an energetic bias is the cause of the increased correlation strength and reduction of scatter observed in well-sampled regular LGRBs, but we encourage future observing strategies that provide increased sampling for less energetic events nonetheless.

We test for further biases in well-sampled regular LGRBs by examining the distributions of two additional properties, the observed peak flux and redshift, in Appendix \ref{sec:appendix well sampled lgrb distributions}. We find that well-sampled regular LGRBs are strongly biased towards brighter observed peak fluxes, which is likely the cause of their increased sampling. Additionally, they are biased towards lower redshifts. Since they are not biased in the rest frame brightness (as shown in Figure \ref{fig:lum_dist}), their bias towards lower redshifts accounts for their bias towards higher brightnesses in the observed frame. A bias towards lower redshift is not likely to affect the correlation (see Section \ref{sec:redshift_evolution_full}).

\subsubsection{The effect of light curve morphology on the correlation in well-sampled regular LGRBs}
\label{sec:morphology}
The XRT catalogue reports the light curve morphological classifications using the criteria originally detailed in \cite{eva09}, and later updated as detailed on the XRT catalogue web page, where a light curve is defined as canonical if the best-fitting light curve model contains at least two breaks including a flattening break, with $\Delta\alpha\leq-0.3$, and a later steepening break, with $\Delta\alpha\geq0.3$\footnote{The criteria originally detailed in \cite{eva09} required that $\Delta\alpha\leq-0.5$ and $\Delta\alpha\geq0.5$, but these thresholds were relaxed to 0.3 in more recent works as stated on the XRT catalogue web page.}. We highlight that this criteria does not include any requirement for canonical light curves having a final segment with $\alpha \gtrsim 2$ (as is the case illustrated in Figure \ref{fig:canonical}). Using the classifications reported in the XRT catalogue, the majority ($\simeq70$ per cent) of well-sampled regular LGRBs have canonical light curves, whereas the full sample of regular LGRBs (N=382) have a more even mix (with $\simeq47$ per cent canonical). 

As most well-sampled regular LGRBs have a canonical morphology, it is unclear if the increased correlation strength and reduced scatter observed in well-sampled regular LGRBs (see Section \ref{sec:well-sampled}) is due to the improved sampling of the light curves, or their canonical morphologies. Therefore, we examine the effect of light curve morphology on the correlation in detail in Appendix \ref{sec:appendix morphology}. In summary, we separate well-sampled regular LGRBs into two groups based on their light curve morphology: canonical and non-canonical, and test for the correlation in each group. We find the correlation strength is similar in both groups, indicating that morphological type does not drive the increase in correlation strength, and the reduction in scatter, observed in the well-sampled LGRB sample.

The majority of well-sampled regular LGRBs having canonical light curves could be an artefact of different observing strategies, such as different morphological types being linked to different observing cadences, or it could simply be due to chance. A deeper analysis in a future study, comparing their observed frame properties, may be useful in providing further insight.

\subsubsection{Redshift selection effect}
\label{sec:redshift_evolution_full}
It was shown in previous studies, using a partial Spearman's rank test, that the correlation between \lum and \decay is not due to the influence of redshift on either parameter \citep[in multiple wavebands, see][]{oates12,oates15,Racusin16,Hinds23,shilling25}. However, high redshift GRBs are biased towards a narrower range of higher luminosities due to the Malmquist bias \citep[see][for a review]{tee97}. The truncated luminosity distribution in high redshift GRBs may, in turn, affect the correlation observed in high redshift GRBs. Furthermore, GRBs may evolve with redshift \cite[see][and references therein]{llo20} which could also affect the correlation in high redshift GRBs.

\begin{center}
\begin{figure}
\includegraphics[width=\columnwidth]{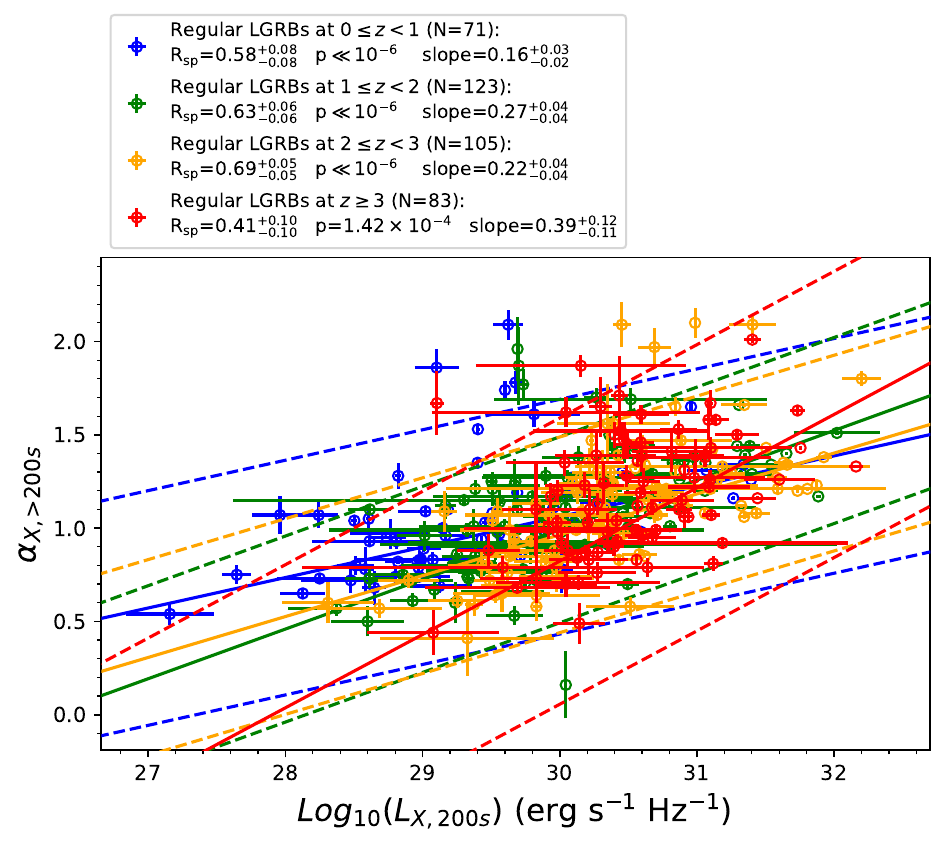}
    \caption{A comparison between the correlation in regular LGRBs separated into different redshift groups. A bias towards higher luminosities is visible in the highest redshift group, which appears to decrease the robustness of the correlation in comparison to the lower redshift groups. This is likely a consequence of the Malmquist bias, narrowing and increasing the luminosity distribution. As our subsamples are not dominated by GRBs at high redshift ($z\geq3$), we do not expect the main conclusions of this paper to be affected by this bias.}
   \label{fig:redshift_evolution_figure}
\end{figure}
\end{center}

In order to test how the correlation is affected by GRBs at different redshifts, we separate regular LGRBs (N=382) into redshift groups and test for the correlation in each as shown in Figure \ref{fig:redshift_evolution_figure}. A bias towards higher luminosities is visible in the highest redshift group, as expected from the Malmquist bias. The $R\mathrm{_{sp}}$ values, and the linear regression slopes, in each redshift group are consistent with those observed in the full regular LGRB sample (N=382) at $\leq2\sigma$, indicating the correlation results are not discrepant as such. However, in the highest redshift group, the fractional (and absolute) uncertainties on these values are largest, the strength of the correlation is the weakest, and the slope is the steepest. These factors indicate that, whilst the strength and the behaviour of the correlation do not become significantly discrepant (at $\geq2\sigma$), the correlation does have decreased robustness at high redshift ($z\geq3$).

In Appendix \ref{sec:redshift_groups_luminosity_cut}, we show evidence that the decreased robustness of the correlation observed at high redshift can be explained by the luminosity selection effect from the Malmquist bias. However, we cannot rule out that it may instead be due to the correlation physically evolving with redshift. This could potentially be achieved by observing fainter GRBs at high redshift using more sensitive instruments. Regardless, we do not expect the main conclusions of this paper (derived from analyses of the full redshift range) to be affected by any such redshift dependency as our subsamples are not dominated by high redshift ($z\geq3$) GRBs.

\section{DISCUSSION}
\label{sec:discussion}

\begin{table*}
    \renewcommand{\arraystretch}{1.4}
	\centering
	\caption{Comparison of the luminosity--decay correlation results for X-ray afterglows in each subsample and in \protect\cite{Racusin16}. Col. (1): The parent sample that subsamples are drawn from. Col. (2): The subsamples under each parent sample. Col. (3): Number of GRBs in each subsample. Cols. (4-6): Spearman's rank correlation coefficient and null hypothesis and corresponding $\geq3\sigma$ confidence interval flag (Y = yes, N = no). Cols. (7-8): Linear regression analysis slope and intercept.}
	\label{tab:summary}
	\begin{tabular}{llc c ccc c cc}
	\hline
     & &  & & \multicolumn{3}{|c|}{Spearman's rank} & & \multicolumn{2}{|c|}{Linear regression} \\

     \cline{5-7} 
     \cline{9-10} 

    {Sample} & {Subsample}  & {Number of GRBs} & & {Coefficient} & {Null hypothesis} & {$\geq3\sigma$ flag} & & {Slope} & {Intercept}\\
    (1) & (2) & (3) & & (4) & (5) & (6) & & (7) & (8) \\

	\hline
 
    \multirow{2}{*}{Full Sample}
     & \textit{Full sample$^{a}$ \protect\citep{Racusin16}} & 246 & & $0.44$ & $\ll 10^{-6}$ & Y & & $0.29^{+0.03}_{-0.06}$ & $-7.47^{+1.80}_{-0.98}$ \\
     & Full sample & {427} & & ${0.54 \pm 0.04}$ & $\ll 10^{-6}$ & Y & & ${0.18^{+0.02}_{-0.02}}$ & $-4.37^{+0.47}_{-0.47}$ \\
    \hline
    
    \multirow{5}{*}{LGRBs}
        & \textit{All LGRBs \protect\citep{Racusin16}} & 237 & & $0.59$ & $\ll 10^{-6}$ & Y & & $0.27^{+0.04}_{-0.04}$ & $-6.99^{+1.23}_{-1.11}$ \\ 
        & All LGRBs &{395} & & ${0.58 \pm 0.04}$ &  $\ll 10^{-6}$  & Y & & $0.19^{+0.02}_{-0.02}$ & $-4.55^{+0.52}_{-0.52}$ \\ 
        &{Low-luminosity LGRBs} &{13} & & $0.24 \pm 0.26 $ & $>0.10$  & N & & $0.18^{+0.11}_{-0.11}$ & $-4.17^{+3.14}_{-3.02}$ \\ 
        & Regular LGRBs &{382} & & ${0.58 \pm 0.04}$ & $\ll 10^{-6}$ & Y & & $0.21^{+0.02}_{-0.02}$ & $-5.35^{+0.60}_{-0.60}$ \\ 
    & {well-sampled regular LGRBs} &{102} & & ${0.80 \pm 0.04}$ & ${\ll 10^{-6}}$  & Y & & $0.21^{+0.03}_{-0.03}$ & $-5.20^{+0.96}_{-0.94}$ \\ 
    \hline
    
    \multirow{4}{*}{SGRBs}
    & \textit{All SGRBs \protect\citep{Racusin16}} & 9 & & $-0.07$ & $>0.10$ & N & & $0.16^{+0.10}_{-0.39}$  & $-3.40^{+11.16}_{-2.77}$  \\ 
    & All SGRBs & {32} & & ${0.29 \pm 0.20}$ & $>0.10$ & N & & $0.13^{+0.08}_{-0.08}$  & $-2.66^{+2.87}_{-2.21}$  \\ 
    & SGRBs with EE & {8} & & ${-0.18 \pm 0.45}$ & $>0.10$ & N & &  $-0.10^{+0.52}_{-0.05}$ &  $-3.79^{+2.29}_{-13.99}$ \\
    & Regular SGRBs & {24} & & ${0.39 \pm 0.23}$ & $5.72\times10^{-2}$ & N & & $0.12^{+0.05}_{-0.04}$  & $-2.51^{+1.19}_{-1.33}$  \\ 
	\hline
	\end{tabular}
    \begin{minipage}[c]{2\columnwidth}
    {$^{a}$The full sample in \protect\cite{Racusin16} does not contain morphological corrections for prompt emission contamination of the afterglow.}
 \end{minipage}
\end{table*}

\subsection{Comparison of results}
The luminosity--decay correlation has previously been observed in \swift/XRT light curves as reported in \cite{oates15,Racusin16}, as well as being observed in the radio, optical/UV, X-ray and GeV wavebands \citep{shilling25,oates09,oates15,Racusin16,Hinds23}. In \cite{shilling25}, the correlation was compared between each waveband highlighting that the strengths of the correlation in each waveband are consistent within $1\sigma$ and are all observed at a significance level of $\geq3\sigma$. In this study, we have re-analysed the correlation in the X-ray using a larger sample of afterglows, which we split into various subsamples within this waveband. Table \ref{tab:summary} summarizes the correlation results between each subsample examined in this study, and for the sample examined in \cite{Racusin16}. 

\subsubsection{Comparison with previous results}
First, we compare the correlation results from this study with those presented in \cite{Racusin16} for each sample in common (full sample, LGRBs and SGRBs). The correlation results in the full samples are not directly comparable as morphological corrections had not been applied to the full sample in \cite{Racusin16}. Both studies do not observe the correlation in SGRBs, although with our larger sample we recover a stronger Spearman's rank. The Spearman's rank for the LGRB sample in this study (N=395) is consistent with that for the LGRB sample in the previous study (N=237) within $1\sigma$, and their linear regression slopes are consistent within $2\sigma$. The slightly different slopes could be due to statistical fluctuation arising from the largely different sample sizes, different linear regression routines, or different best-fit models.

We briefly investigate the cause of the slightly different linear regression slopes measured in each study by analysing the 237 values of \lum and \decay measured in \cite{Racusin16} for a correlation using our correlation analysis and linear regression fitting routines. We calculate a Spearman's rank coefficient of $0.58 \pm 0.05$ and a slope of $0.24 \pm 0.03$. These values are consistent with those reported in \cite{Racusin16} within $1\sigma$. This indicates that the slight difference in linear regression slope for the LGRB samples in each study (see Table \ref{tab:summary}) is not likely due to the different linear regression routines. Instead, it could be due to statistical fluctuation in the different samples, or due to the slightly different best-fit models in each study; for some GRBs, different best-fit models could lead to differences in the identification of steep decay segments. In turn, these could affect the measurements of \lum, due to the different extrapolations to $t_\mathrm{rest}=200$ s, and of \decay, due to the exclusion of different data when fitting a single power-law.

\subsubsection{Comparison between subsamples}
Next, we compare the results between each subsample in this study. We find that no correlation, at a confidence interval of $\geq3\sigma$, is recovered in any SGRB subsample. However, the correlation strength increases and the p-value decreases in regular SGRBs, after excluding SGRBs with extended emission. We examine this further in Section \ref{sec:short_simulation}. As for LGRBs, we observe the correlation in the full sample of LGRBs as well as the subsamples of regular LGRBs and well-sampled regular LGRBs, but not in low-luminosity LGRBs. Furthermore, low-luminosity LGRBs appear to decay on average more quickly than expected from the LGRB population as shown by their offset from the regular LGRB best-fit linear regression slope (see Figure \ref{fig:long_low_lum}). However, similarly to SGRBs, there is a low number of events (N=13) in the subsample of low-luminosity LGRB meaning it is not currently reliable for interpretation. 

We also see that the correlation strength in our subsample of well-sampled regular LGRBs is significantly increased and that scatter is significantly reduced compared to all other subsamples. This suggests that regular LGRBs fundamentally follow the correlation tightly, and that scatter is introduced if they are not observed consistently over each epoch of their lifetime as \decay may not be measured over a range of temporal behaviours spanning a large portion of the light curve.

\subsection{Origin of the correlation}
The underlying cause of the correlation is not yet fully understood. The correlation has been observed with consistent strength and behaviour across multiple wavebands covering 15 orders of magnitude in frequency, from radio photons to GeV photons. This places the constraint that any possible cause of the correlation must be achromatic such that the correlation is produced with consistent behaviour across all wavebands. As such, previous studies have proposed two candidate mechanisms as possible causes of the correlation: geometric effects due to the jetted emission or some mechanism that regulates the rate at which energy is released by the GRB central engine \citep{oates12,oates15,Racusin16,Hinds23,shilling25}. The results of this paper, that the correlation is observed in an updated sample of LGRB X-ray afterglows, are consistent with these two possible causes. 

The microphysical parameters of a GRB play a key role in dictating the observed spectra and light curves, for instance; the density of the surrounding medium $n$, the magnetic field $\epsilon_B$, and the fraction of shock energy transferred to the electrons $\epsilon_e$ \citep{sar98}. Microphysical parameters are typically assumed to be constant, but time-varying microphysical parameters have previously been invoked in order to explain the observed properties in certain GRB light curves \citep[as required for GRBs 091127; 130427A; and 190114C in][respectively]{fil11,van14,mis21}, and may be an example of a mechanism that can cause the correlation by regulating the energy release. The time-varying microphysics scenario could be constrained as a cause through numerical simulations of GRB afterglows in the framework of the standard fireball model. For instance, microphysical parameters could be set to vary with time in these simulations and then test if this can reproduce the observed correlation \citep[see][]{bos14}. As highlighted in \cite{pan06}, care should be taken to ensure that the observed results - in this case, the correlation - are not reproduced from the simulations in an unrealistic or contrived fashion. For example, the rate of change of a time-evolving parameter should be constrained so that it cannot abruptly change without physical explanation.

The viewing angle is an additional parameter that can affect the observed light curve properties. For example, an event observed off-axis will be fainter, peak later, and have a slower decline rate compared to the same event observed closer to its jet axis \citep{granot02,pan08}. Consequently, the correlation may be due to a range of off-axis viewing angles in a sample. However, jet structures can also affect the observed light curve for a given viewing angle which may add complexity to the viewing angle scenario, particularly if GRBs do not have a universal jet structure. If the correlation is caused by viewing angle effects, we would expect there to be a correlation between the viewing angle and the \decay--\lum plane. Furthermore, we would expect to observe it (albeit potentially with different behaviour) in different populations of GRBs, provided their emission is jetted, such as SGRBs and LGRBs because geometric effects, due to observer viewing angles, occur independently of progenitor or environment. However, as shown in Section \ref{sec:subsamples}, we do not observe the correlation in our SGRB subsample. We note that this does not conclusively rule out the possibility of the correlation in SGRBs or viewing angle as the cause. In Section \ref{sec:short_simulation}, we estimate how many additional SGRBs we need before we can begin to confidently make conclusions about the presence of the correlation in SGRBs.

\subsection{Simulating future regular SGRB samples}
\label{sec:short_simulation}
In Section \ref{sec:outlier} we showed that no correlation is recovered for regular SGRBs at a confidence interval of $\geq3\sigma$. However, after removing SGRBs with extended emission, the $R_\mathrm{sp}$ result increases from $R_\mathrm{sp}=0.29\pm0.20$ in the full SGRB sample to $R_\mathrm{sp}=0.39\pm0.23$ in the regular SGRB subsample. Furthermore, the p-value decreases from $>0.10$ to $5.72\times10^{-2}$ meaning the confidence interval increases, though insufficiently as to provide evidence for a correlation at $\geq3\sigma$. These improved statistics imply that the correlation may exist in regular SGRBs but is not observed simply due to sample size limitations, or (additionally) due to the systematically lower and narrower distribution of \lum in SGRBs compared to LGRBs, with a mean and standard deviation $\log_{10}$(\lum) of $28.52 \pm 0.99$ dex in SGRBs compared to $30.06\pm0.95$ dex in LGRBs, as well as a range of 3.96 dex in SGRBs compared to 7.77 dex in LGRBs. A difference as such is consistent with other studies showing that the afterglows of SGRBs are systematically fainter than LGRBs \citep[see][]{mar13,kan10,nic12}. These properties can be explained following that the brightness of a GRB afterglow depends, in part, on its energetics and the density of the surrounding environment \citep{gra02a} and SGRBs have been observed to explode into lower density environments \citep[see][]{fon15}. 

The lower and narrower distribution of luminosities in SGRBs hinders the significance of the correlation as afterglows covering a broad luminosity range are required to fully capture the behaviour of the correlation (as shown in Appendix \ref{sec:redshift_groups_luminosity_cut}). Assuming the correlation exists, in order to potentially observe the correlation at confidence level of $\geq3\sigma$, one would either: need to detect SGRBs with a wider range in luminosity, or require a larger sample size with the narrower luminosity range. Since the former may not happen, we examine the latter. To investigate this possibility, we assume that the correlation does exist in regular SGRBs and use a Monte Carlo simulation to synthetically increase the regular SGRB sample size based on the measured distributions of \lum and \decay and test for a correlation. This enables an estimate of how many more regular SGRBs we would need in our sample in order to observe the correlation at a significance of $\geq3\sigma$, and in turn, how many more years of \swift/XRT observations this corresponds to. 

For a total of 10 iterations, we add +1 synthetic regular SGRB to our sample of 24 regular SGRBs cumulatively for each iteration. Each iteration is run over $10^{5}$ trials. For each trial, a number of synthetic regular SGRBs (corresponding to the current iteration number) are added by randomly selecting paired measurements of \decay and \lum from our measured results. We also add noise to these measurements by randomly selecting how many standard deviations to add as noise by randomly drawing a single value from a uniform distribution of 100 evenly spaced values between $-3$ and $+3$. This value is added to each measurement as a multiple of the standard deviation of each respective measurement. Using a single standard deviation level for both measurements means that any random increase or decrease in \decay is paired with the same level of change (between $\pm3\sigma$) in \lum in the same direction (increase or decrease). This means that any new brighter SGRBs will tend to decay more quickly than any new fainter SGRBs. This key feature of our simulation is consistent with the behaviour observed in the correlation, since our underlying assumption is that SGRBs follow the correlation and our goal is to calculate how many more SGRBs we need in our sample size before we can robustly observe the correlation under this assumption. For each iteration, we record the number of trials out of $10^{5}$ where a correlation is observed at a significance level of $\geq3\sigma$.

\begin{center}
\begin{figure}
\includegraphics[width=\columnwidth]{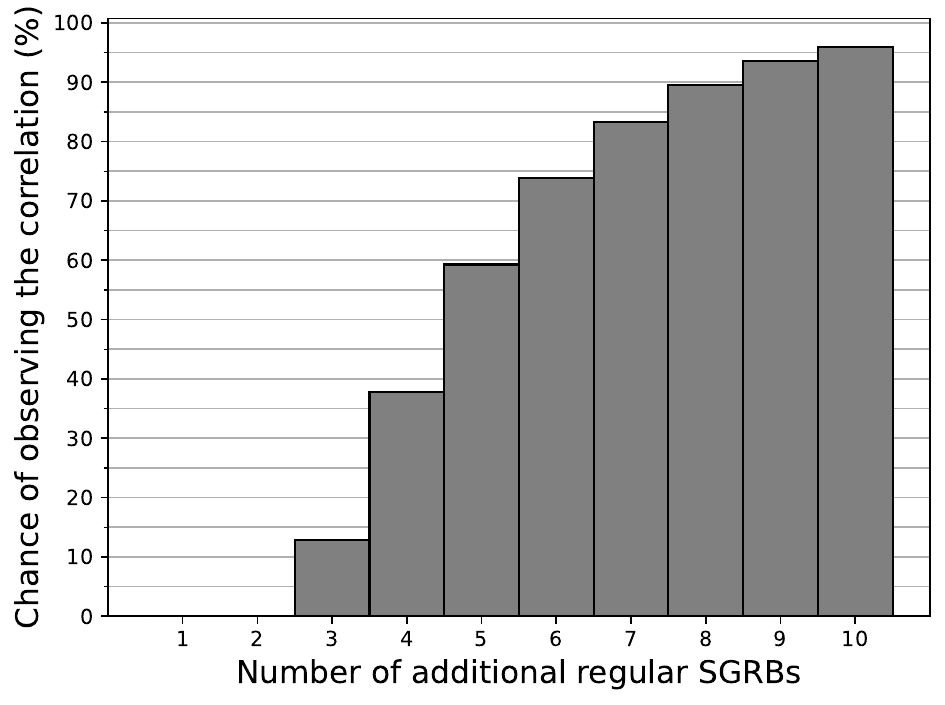
}
    \caption{Estimated chance of observing the correlation at a confidence interval of $\geq3\sigma$ for each larger sample size of regular SGRBs. For each sample size, the chance of observing the correlation is calculated over $10^{5}$ trials. The simulation assumes the correlation exists in regular SGRBs and calculates the probability of observing it for each larger sample size. To obtain a $>90$ per cent chance of observing the correlation, we would require another 9 or more additional SGRBs.}
   \label{fig:sgrb_simulation_results}
\end{figure}
\end{center}

The results of the simulation are shown in Figure \ref{fig:sgrb_simulation_results}, where the predicted probability of observing the correlation for each larger simulated sample size is shown. Regular SGRBs that pass our selection criteria are observed a rate of $\simeq1$ per year. To obtain a $>90\%$ chance of observing the correlation, our simulation suggests we would require another 9 or more SGRBs, which corresponds to another $\gtrsim9$ years of observing with \swift/XRT if there is no change in GRB follow-up criteria or strategy. If the correlation is not observed with 9 or more additional regular SGRBs, we would need to sample a wider distribution of luminosities before we can rule out the existence of the correlation in SGRBs.

\subsection{Applications of the correlation}
We discuss the applications of the correlation, including the measurement of the luminosity and the average rate of decay at a later time for GRBs that are not observed close to 200 s in the rest frame. Also, the use of the correlation to investigate the potentially different populations of radio--loud and radio--quiet GRBs. For these applications, we use our subsample of regular LGRBs with well-sampled light curves (N=102), as the correlation is observed in this subsample with the highest Spearman's rank and the least scatter out of all subsamples, and has a slope that is consistent with the parent LGRB sample.

\begin{center}
\begin{figure}
\includegraphics[width=\columnwidth]{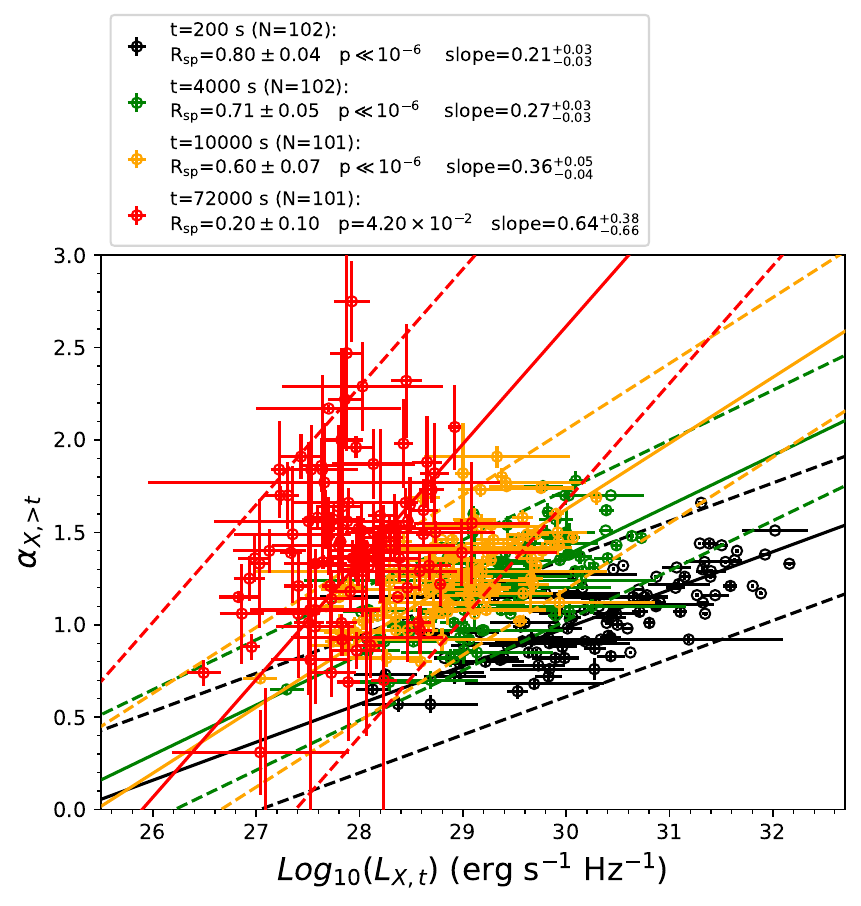}
    \caption{The correlation in well-sampled regular LGRBs (as in Figure \ref{fig:lgrbs_by_data_quality}) with the luminosity measured at different times, and the average rates of decay measured from each respective time. We re-analyse the light curves at each later time and exclude measurements with large errors, as in Section \ref{sec:measurements}, meaning that the sample size may change. The correlation weakens with each later time but remains statistically significant at $\geq3\sigma$ up to $10$ ks.}
   \label{fig:later_time_correlation}
\end{figure}
\end{center}

\subsubsection{GRBs observed later than 200 s of their trigger time in the rest frame}
\label{sec:later times}
There may be GRBs observed by \swift/XRT later than 200 s in the rest frame after their initial detection in the $\gamma$-ray. This is particularly true for \swift/XRT target-of-opportunity observations of GRBs detected by other instruments. For example, the Einstein Probe \citep[EP][]{yua22} and SVOM \citep{att22} have exceptionally sensitive detectors in the soft and hard X-ray bands respectively. As such, each instrument is effective at detecting GRBs which are systematically faint either due to being a high redshift source or having a largely off-axis viewing angle \citep[see][]{Wei25}. These objects could increase the number of GRBs in the low-luminosity region of the correlation and verify whether or not these decay on average more slowly, as expected from the correlation. If such objects are to be included in the correlation analysis, then it may be necessary to test for the correlation using a later time instead of 200 s in the rest frame. As such, we examine how the luminosity--decay correlation is affected by the measurement time.

We re-measure the luminosity and the average rate of decay in our well-sampled regular LGRBs (N=102) following the same method as in Section \ref{sec:measurements} but instead using later times of 4 ks, 10 ks, and 72 ks. For each later time, we exclude any measurements with large uncertainties of $>2$ dex for the luminosity or $>0.25$ for the average rate of decay as in Section \ref{sec:measurements}, meaning that the sample size may change between each version. Figure \ref{fig:later_time_correlation} shows a plot of the luminosity and the average rate of decay measured using each later time. The correlation is observed with a statistical significance of $\geq3\sigma$ up to the time of 10 ks. By $72$ ks, the significance decreases to $<3\sigma$. The strength of the correlation decreases with each later successive time and the slope increases. Similar conclusions are found when testing this in the full sample of GRBs (N=427) instead of in well-sampled regular LGRBs (N=102) as shown in Appendix \ref{sec:later_time_full}.

\subsubsection{Application to potentially different radio subclasses}
Based on the findings of \cite{lloyd17,lloyd19} determining that radio--quiet LGRBs have shorter intrinsic (rest frame) prompt durations compared to radio--loud LGRBs, it has been suggested that radio--loud and --quiet LGRBs are different populations entirely. A progenitor-based explanation is discussed in \cite{zha21,lloyd22,chakra23}. However, \cite{osb21} studied these differences between radio--loud and --quiet LGRBs and established that these differences could be a result of selection effects or sample incompleteness. 

The luminosity--decay correlation may be used as a means of investigating whether or not radio--loud and --quiet GRBs belong to different populations with different progenitor systems. This was previously considered in \cite{shilling25}, however without a sample of radio--quiet afterglows to compare against a radio--loud sample, this was not possible. Therefore, we revisit this application of the correlation here by selecting radio--loud and --quiet GRBs from our sample of X-ray afterglows depending on the detection of their radio counterparts. 

An intrinsically radio--loud LGRB could be misclassified as radio--quiet if its radio afterglow is too faint to be detected. This potential source of contamination is highlighted in \cite{lloyd17,lloyd19}, where an attempt to reduce such contamination is made by selecting LGRBs with large $E_\mathrm{iso,\gamma}$ values. Although this introduces an energetic bias, $\mathrm{log_{10}}(E_\mathrm{iso,\gamma})$ is correlated with radio detectability \citep[see][]{Chandrafrail12} which, in turn, means that selecting bright GRBs increases the likelihood of a radio afterglow being detectable. As a result, any energetic GRB with radio follow-up but without a radio detection is more likely to be intrinsically radio--quiet.

We follow this method and select bright GRBs from our subsample of well-sampled regular LGRBs (N=102). We use the same threshold as in \cite{lloyd17,lloyd19} of $\mathrm{log_{10}}(E_\mathrm{iso,\gamma}) \geq 52$ erg, resulting in a sample of 97 GRBs. As this selection only excludes 5 out of 102 well-sampled regular LGRBs, we do not expect the energetic bias from this selection to affect the correlation results. We use the radio catalogue from \cite{shilling25} and update this with radio observations (at any frequency) from the literature to up until November 2024 (the same as in Section \ref{sec:initial sample}) resulting in 502 GRBs with radio observations covering a frequency range of $0.2-667$ GHz. We use this catalogue to select bright well-sampled regular LGRBs (N=97) that have radio follow-up, of which there are N=41. We then select those with at least one $\geq3\sigma$ detection at any frequency for our radio--loud sample (N=24) and those without for our radio--quiet sample (N=17).

\begin{center}
\begin{figure}
\includegraphics[width=\columnwidth]{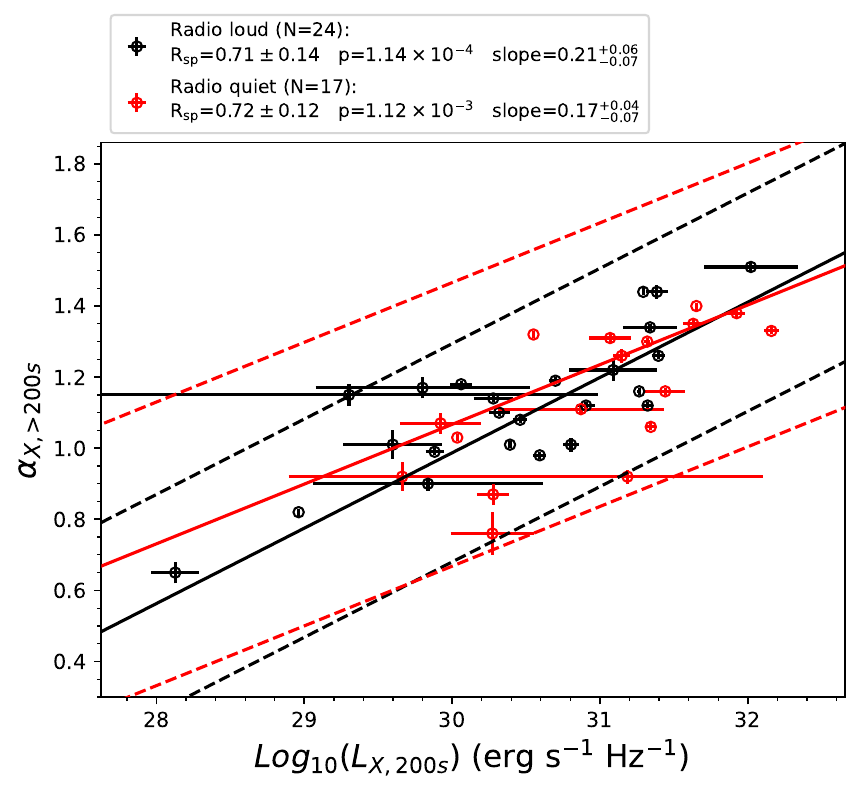}
    \caption{\decay against \lum for bright well-sampled LGRBs which are radio-loud (detected in the radio at $\geq3\sigma$) and radio-quiet (observed in the radio but not detected at $\geq3\sigma$). The consistent results between each population and a similar distribution of scatter in the \decay--\lum plane between the two subclasses suggest that radio-loud and -quiet GRBs are not physically different.}
   \label{fig:radio_samples}
\end{figure}
\end{center}

Figure \ref{fig:radio_samples} shows a plot of \decay against \lum for the radio--loud (N=24) and --quiet (N=17) well-sampled regular LGRBs. The radio--loud and --quiet samples have Spearman's ranks of $0.71 \pm 0.14$ and $0.72 \pm 0.12$ respectively both with confidence intervals of $\geq3\sigma$, as well as linear regression slopes of $0.21^{+0.06}_{-0.07}$ and $0.17^{+0.04}_{-0.07}$ respectively. Our results show that the correlation is consistent between radio--loud and --quiet LGRBs which suggests they may not be physically different classes, which is in agreement with the argument of \cite{osb21}.

\section{Conclusions}
\label{sec:conclusions}
The luminosity--decay correlation has previously been observed in LGRBs across multiple frequencies, from radio to GeV, including the X-ray using a sample of afterglows observed with \swift/XRT over the first 10 years of operation. We have compiled an updated sample of X-ray afterglows observed with \swift/XRT over 20 years, resulting in a sample size increase from N=246 in the previous X-ray study to N=427. Excluding steep decay segments and flare intervals in any light curves, we measured the luminosity at $200$ s in the rest frame, \lum, and the average rate of rate of decay from $200$ s in the rest frame, \decay, using a single power-law fit. We observed a statistically significant correlation between \lum and \decay, consistent with previous results, showing that GRB X-ray afterglows which are initially bright tend to decay on average more quickly. 

We separated GRBs into various subsamples for individual analyses. First, we separated GRBs into two classes of long ($T_{90}>2$ s) and short ($T_{90}\leq2$ s). We then further separated LGRBs into low-luminosity LGRBs ($E_{iso,\gamma}<10^{51}$ erg) and regular LGRBs ($E_{iso,\gamma}\geq10^{51}$ erg), as well as further separating SGRBs into SGRBs with extended emission (with low-level emission after the main pulse) and regular SGRBs (without extended emission) using a list compiled from the literature. We also selected a subsample of regular LGRBs that have well-sampled light curves (with $\geq3$ data points in $\geq5$ dex time bins). There is evidence of a statistically significant correlation in the subsamples of: all LGRBs, regular LGRBs, and regular LGRBs with well-sampled light curves. Furthermore, in the subsample of regular LGRBs with well-sampled light curves, the strength of the correlation increases to $0.80\pm0.04$ and the degree of scatter is reduced, whilst the slope of $0.21\pm0.03$ remains consistent with the parent sample of regular LGRBs. This suggests all regular LGRBs follow the correlation tightly, and that scatter is introduced if they are not observed consistently over each epoch of their lifetime. No evidence of a correlation at a significance of $\geq3\sigma$ is observed in low-luminosity LGRBs or any SGRB subsample.

We discussed the possible causes of the correlation, namely geometric effects due to the jetted emission, such as the distribution of observer angles with respect to the jet axis, or some mechanism that regulates the rate at which energy is released by the GRB, such as time-varying microphysical parameters. If the correlation is due to viewing angle, we might expect to observe the correlation across different populations of GRBs regardless of their different progenitors or environments, provided their emission is jetted and produced via the same mechanisms. These factors suggest it may be possible to constrain the cause of the correlation by comparing the correlation in two subsamples that fit this criteria, such as (regular) SGRBs and (regular) LGRBs. However, despite being on the borderline of statistical significance with a p-value of $5.72\times10^{-2}$, there is insufficient evidence of a statistically significant correlation (at $\geq3\sigma$) in regular SGRBs and therefore our results do not constrain the possible cause.

We re-measured the luminosity and average rate of decay for our subsample of regular LGRBs with well-sampled light curves at later times of 4 ks, 10 ks, and 72 ks to test how late follow-up observations can commence before the correlation can no longer be recovered. Our results showed that the correlation is stronger at earlier times, but a statistically significant correlation can be recovered using later times up to 10 ks. As such, GRBs that are observed late can be included in the correlation analyses, although earlier times closer to 200 s in the rest frame are best. We also use the correlation to examine the possibility of two separate populations of radio--loud and --quiet LGRBs. We selected radio--loud and --quiet LGRBs from our subsample of regular LGRBs with well-sampled light curves and examined the correlation in each for comparison. We observed a significant correlation in each population with a strength and slope that are consistent with each other within $1\sigma$. These results suggest it is unlikely that they are different populations.

Our subsample of regular LGRBs with well-sampled light curves contains the strongest correlation in addition to the least scatter. As a result, we encourage follow-up observations of future GRBs with good sampling up to the time when the afterglow brightness decays below the instrument sensitivity limit. We particularly encourage this follow-up strategy for SGRBs so that the correlation in SGRBs can be observed, or ruled out, with more confidence.

\section*{Acknowledgements}
This work made use of data supplied by the UK \swift Science Data Centre at the University of Leicester. SPRS acknowledges support from an STFC PhD studentship, the
Faculty of Science and Technology at Lancaster University, and NASA under award number 80GSFC21M0002. RG is sponsored by the National Aeronautics and Space Administration (NASA) through a contract with ORAU. GPL and CT are supported by a Royal Society Dorothy Hodgkin Fellowship (grant Nos. DHF-R1-221175 and DHF-ERE-221005).

\section*{Data Availability}
All light curve data are available online as referenced. Table \ref{tab:measurements} is available in its entirety online as supplementary material from the journal website, and from CDS via anonymous FTP at \texttt{cdsarc.u-strasbg.fr} (130.79.128.5), or from the CDS catalogue at \url{https://cdsarc.cds.unistra.fr/viz-bin/cat/J/MNRAS/Vol/Article}.

\bibliographystyle{mnras}
\bibliography{GRB} 

\appendix

\section{Light curves with misidentified flare intervals}
\label{sec:appendix extra light curves}
\begin{center}
\begin{figure*}
\includegraphics[width=0.85\columnwidth]{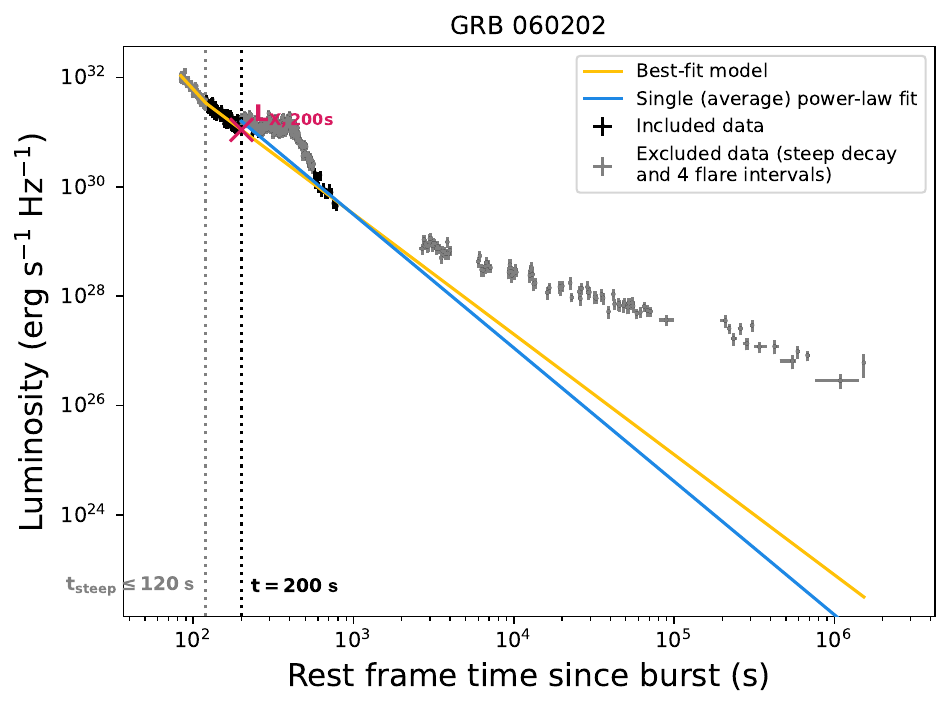}
\includegraphics[width=0.85\columnwidth]{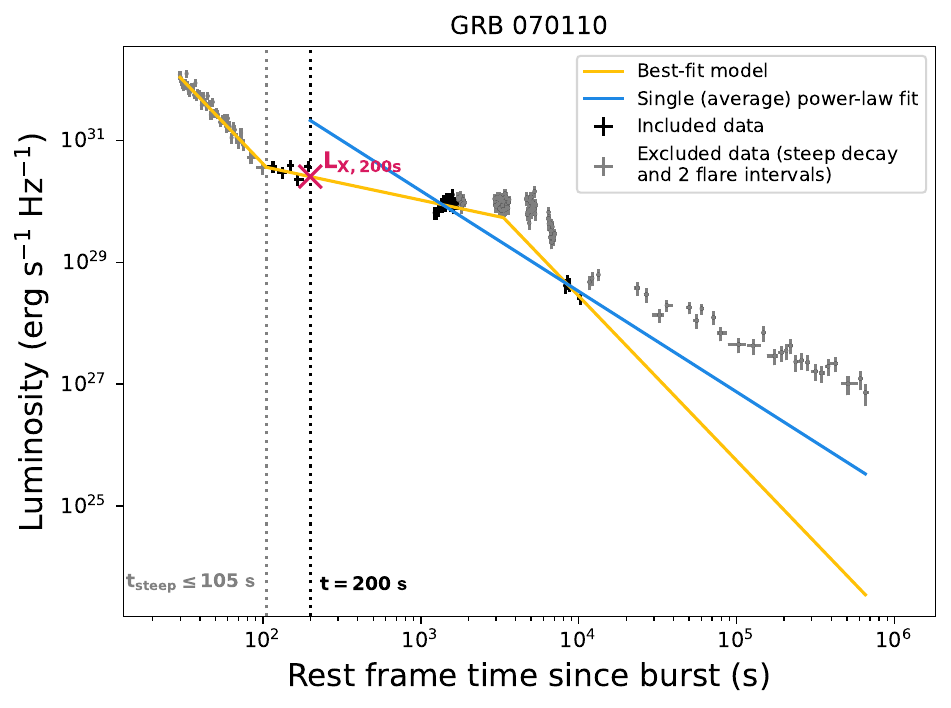}
\includegraphics[width=0.85\columnwidth]{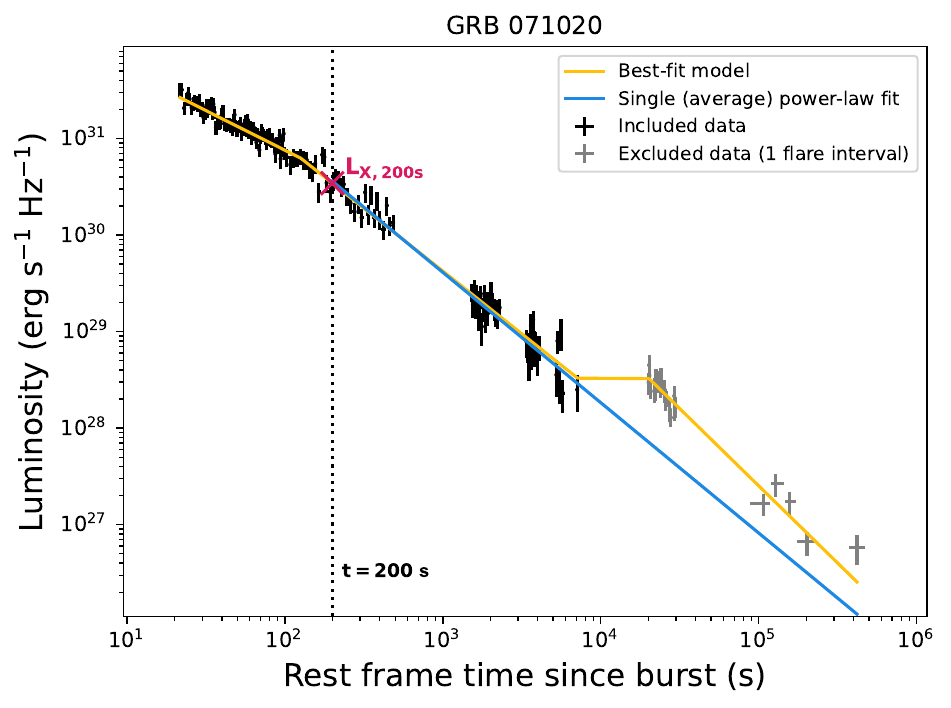}
\includegraphics[width=0.85\columnwidth]{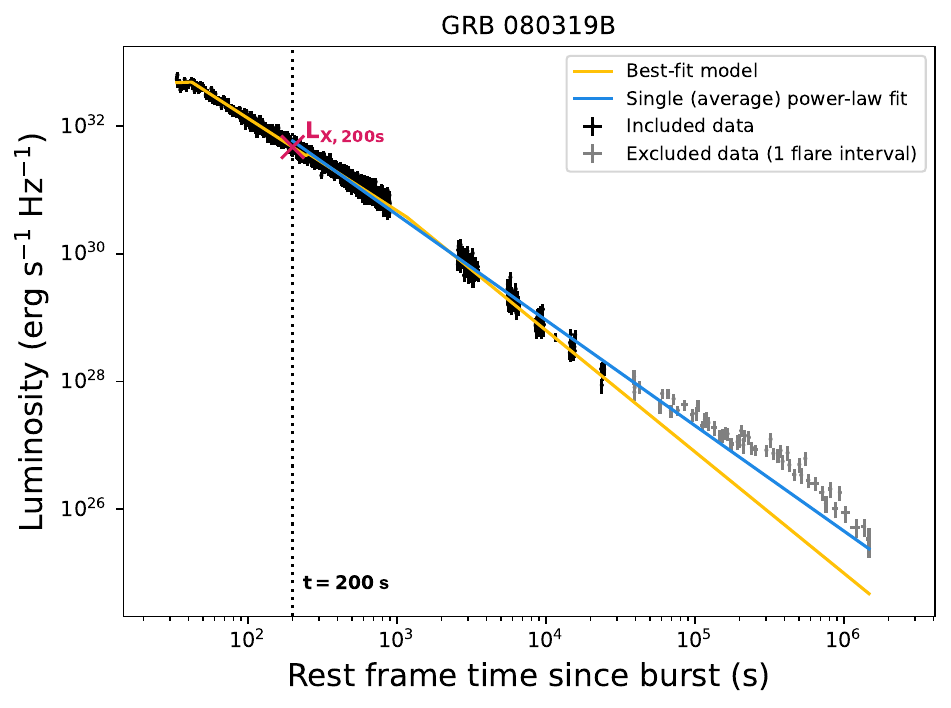}
\includegraphics[width=0.85\columnwidth]{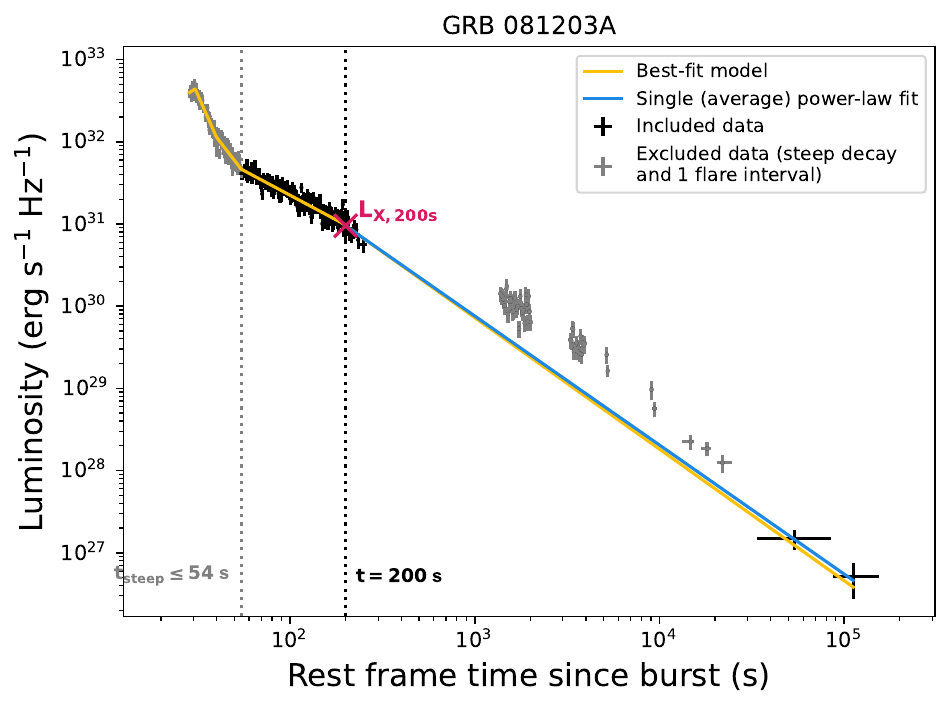}
\includegraphics[width=0.85\columnwidth]{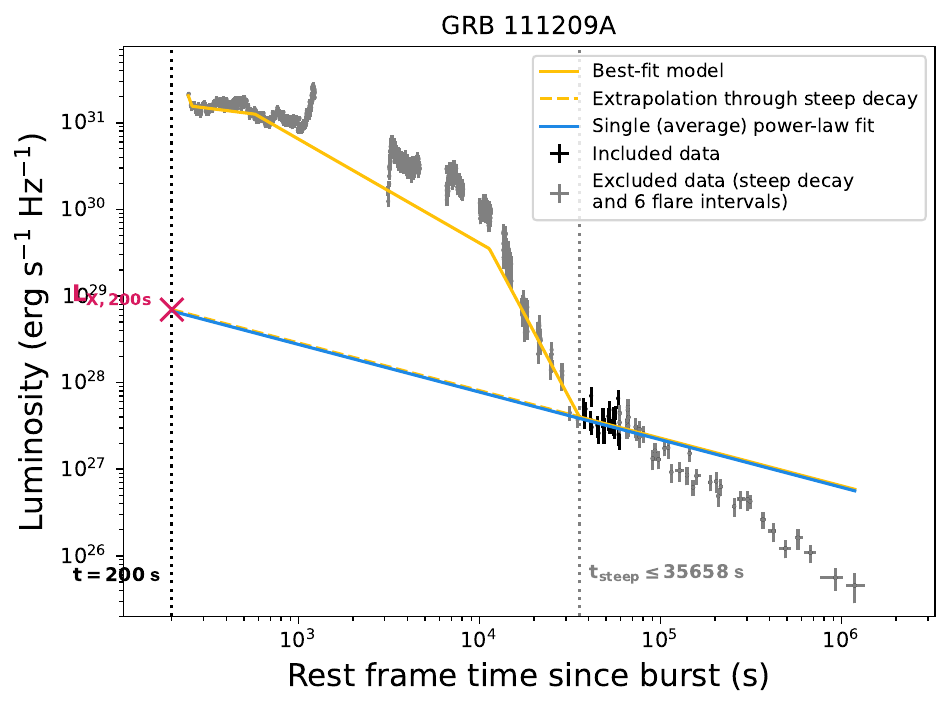}
\includegraphics[width=0.85\columnwidth]{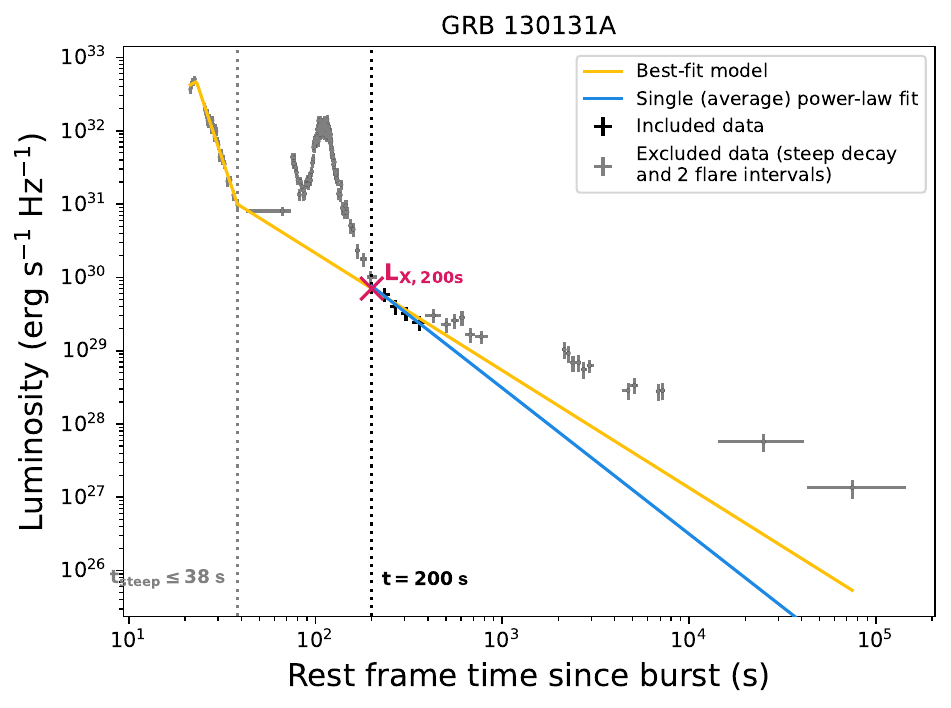}
\includegraphics[width=0.85\columnwidth]{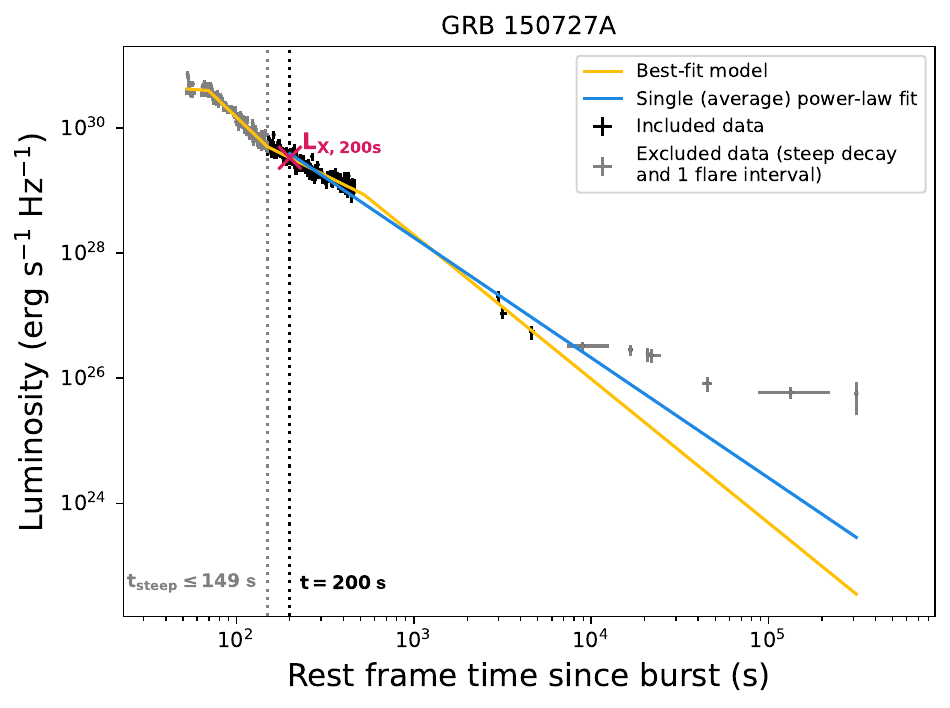}
    \caption{The light curves of 8 GRBs that were excluded from our sample in Section \ref{sec:measurements} due to potentially misidentified flares. The best-fit models models and the single power-laws are overlaid where the data within each flare interval has been excluded, demonstrating the misconstruction of the true underlying light curve.}
   \label{fig:bad_flare_removals}
\end{figure*}
\end{center}

There are 8 GRBs which we flag as having misidentified flares that we remove from our sample as mentioned in Section \ref{sec:measurements}. These are GRBs 060202, 070110, 071020, 080319B, 081203A, 111209A, 130131A and 150727A. The light curves for these GRBs are shown in Figure \ref{fig:bad_flare_removals}.

\section{K-correction of the isotropic-equivalent gamma-ray emission}
\label{sec:appendix k-corr}

We use Equation \ref{eqn:eiso} to estimate the isotropic equivalent gamma-ray energy, $E_{iso,\gamma}$, in the co-moving `bolometric' bandpass of $1-10,000$ keV. Using the method in \cite{blo01}, this includes a $k$-correction factor defined as:
\begin{equation}
    k=\frac{S_{\left[\frac{E_1}{(1+z)},\frac{E_2}{(1+z)}\right]}}{S_{[e_1,e_2]}},
	\label{eqn:k-ratio}
\end{equation}
where $S$ is the fluence in units of $\mathrm{keV\ cm^{-2}}$ for a given energy band, $[e_1,e_2]$ is the detector bandpass in which the observed (redshifted) fluence is measured (corresponding to $e_1=15$ keV and $e_2=150$ keV for \swift/BAT), and $[E_1,E_2]$ is the desired energy range in the GRB co-moving frame (corresponding to $E_1=1$ keV and $E_2=10,000$ keV for a `bolometric' range). The fluence in a given bandpass is calculated using: 
\begin{equation}
    S_{[e_1,e_2]}=t_{exp}\cdot\int^{e_2}_{e_1} E\cdot \phi(E) \cdot dE,
	\label{eqn:fluence}
\end{equation}
where $t_{exp}$ is the exposure time of the BAT spectrum, and $\phi(E)$ is the form of the best-fitting model to the BAT spectrum. The model can take the form of a single power-law (PL; see Equation \ref{eqn:simple}), or a power law with an exponential cutoff (CPL; see Equation \ref{eqn:cutoff}):
\begin{equation}
    \phi(E)=K_{50}^{PL} \left( \frac{E}{50 \ \mathrm{keV}}\right)^{\alpha_{PL}},
	\label{eqn:simple}
\end{equation}
\begin{equation}
    \phi(E)=K_{50}^{CPL} \left( \frac{E}{50\ \mathrm{keV}}\right)^{\alpha_{CPL}} \cdot e^{\frac{-E(2+\alpha_{CPL})}{E_P}},
	\label{eqn:cutoff}
\end{equation}
where $E$ is the photon energy in units of keV, $K_{50}^{PL,CPL}$ is the normalization for the PL or the CPL models in units of photons cm$^{-2}$ s$^{-1}$ keV$^{-1}$, $\alpha_{PL,CPL}$ is the dimensionless photon index of each model, and $E_p$ is the peak energy in the cutoff power law model in units of keV. For each GRB, we use the best-fitting model made to the time-averaged spectra created from photons in the $T_{100}$ range as reported in the third BAT catalogue. We retrieve the best-fitting model and the corresponding parameters from the third BAT catalogue. The PL model is the best-fitting spectral model for $\sim95$ per cent of GRBs in our sample, compared to the CPL model which is only best-fitting for $\sim5$ per cent.

\section{Light curve distributions}
\label{sec:appendix extra light curves 2}
\begin{center}
\begin{figure*}
\includegraphics[width=2\columnwidth]{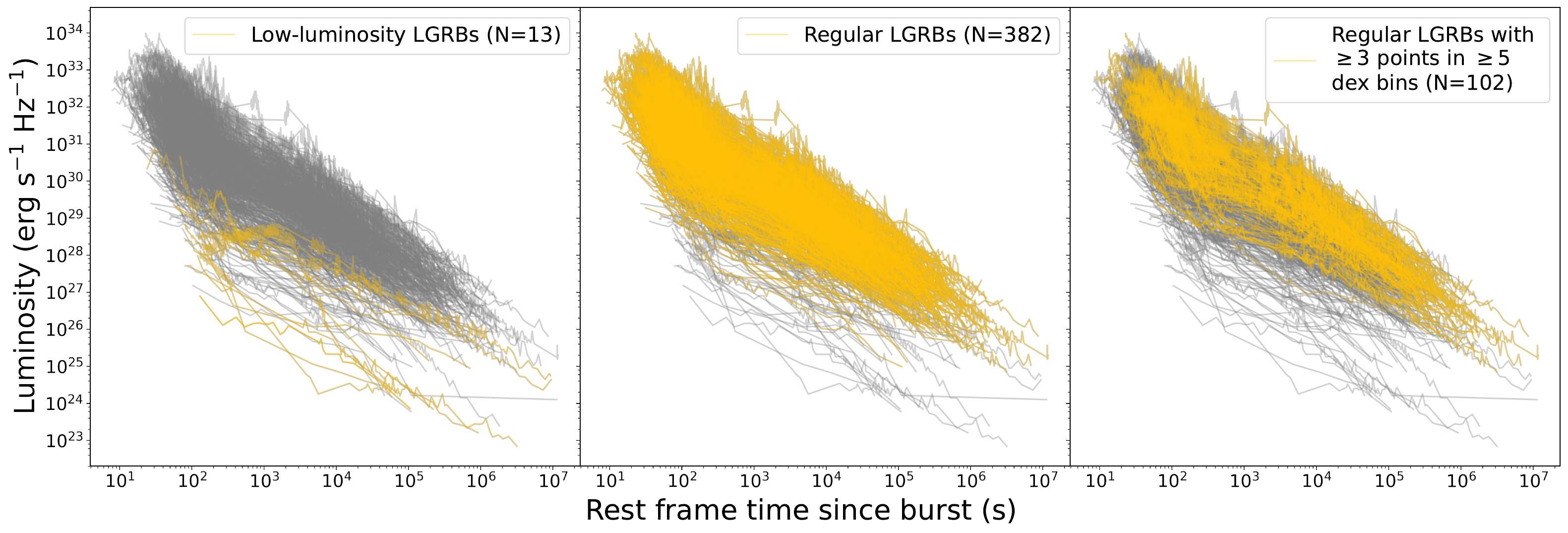}

\includegraphics[width=1.34\columnwidth]{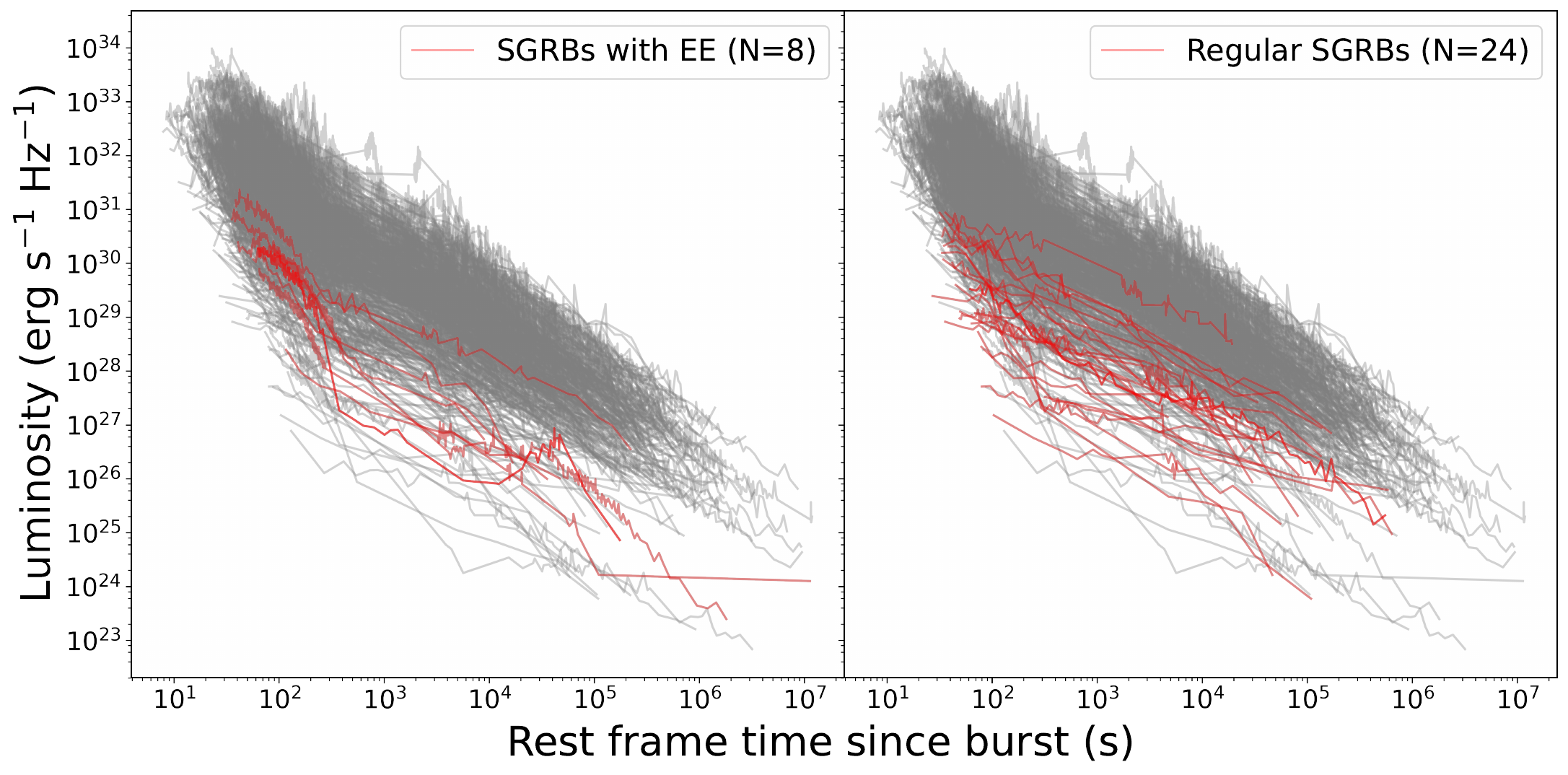}

    \caption{Plots of the light curve distributions for each GRB subsample overlaid onto that of the full sample. Long GRB subsamples are shown in yellow in the top row and short GRB subsamples are shown in red in the bottom row. The full sample is shown in grey in each plot. The convergence of the light curve distributions at later times is consistent with the correlation and is most noticeable in the well-sampled regular LGRB subsample (rightmost plot in the top row) where the correlation is the strongest.}
   \label{fig:long_lc_dist}
\end{figure*}
\end{center}

In Section \ref{sec:subsamples} we separated GRBs into subsamples. The rest frame light curve distributions of each subsample are shown, overlaid onto that of the full sample, in Figure \ref{fig:long_lc_dist}.

\section{Alternative criteria for selecting well-sampled regular LGRBs}
\label{sec: appendix well-sampled options}
\begin{center}
\begin{figure*}
\includegraphics[width=1.6\columnwidth]{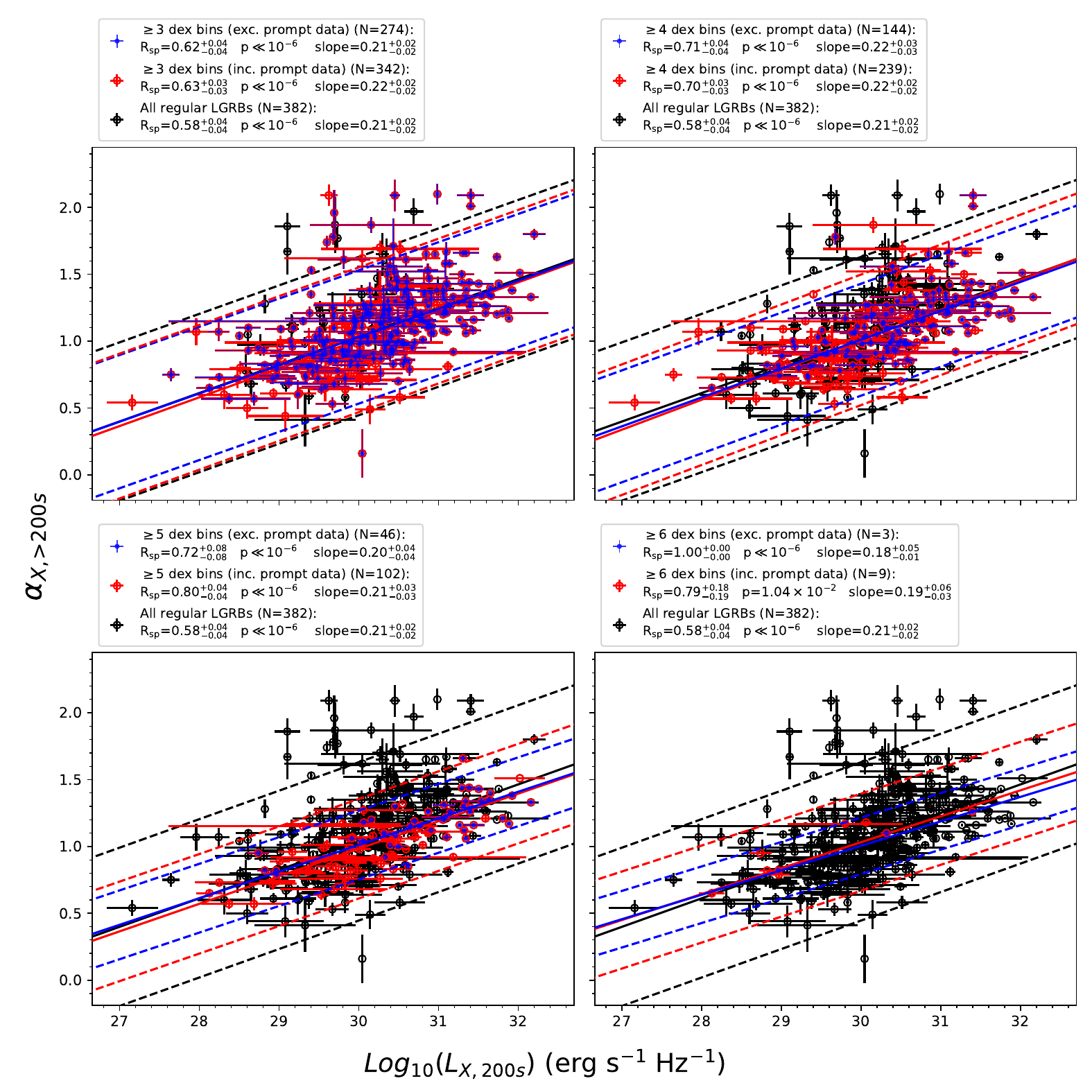}
    \caption{A comparison between the correlation in well-sampled regular LGRBs for different selection criteria of requiring $\geq3$ points in the following number of dex bins: $\geq3$, $\geq4$, $\geq5$, and $\geq6$. Two versions are overlaid onto the parent sample of regular LGRBs: including (`inc.') or excluding (`exc.') the data within flare intervals and steep decay segments (`prompt data'). The result of $R\mathrm{_{sp}}=1.00\pm0.00$ when excluding prompt data and counting over $\geq6$ dex bins is not reliable and has an uncertainty of zero (to 2 decimal places) due to a sample size of only three events. With each successive dex bin requirement, in both versions, the level of scatter decreases, the strength of the correlation increases, whilst the linear regression slope remains consistent. Additionally, for a given dex bin, the sample size is larger when prompt data are included. We therefore opt to include these data in order to maximize the sample size. Similarly, we choose to require good sampling ($\geq3$ points) in $\geq5$ dex bins in order to optimize the sample size with the improved correlation results.}
   \label{fig:well_sampled_criteria}
\end{figure*}
\end{center}

As discussed in Section \ref{sec:well-sampled}, we selected regular LGRBs which have well-sampled light curves using the criterion of containing $\geq3$ points in $\geq5$ dex bins, including data within flare intervals and steep decay segments. Here, we test other options for the selection criteria and examine how the correlation changes in each resulting sample. A suitable range of options include requiring $\geq3$ points in the following number of dex bins: $\geq3$, $\geq4$, $\geq5$, and $\geq6$. Additionally, there is the choice to exclude data within flare intervals and steep decay segments when counting the number of points in each dex bin. Therefore, in two iterations: including and excluding data within flare intervals and steep decay segments, we re-select light curves that meet the criterion ($\geq3$ data points) in each of these numbers of dex bins and test for the correlation in each resulting sample. These alternative selections are shown in Figure \ref{fig:well_sampled_criteria} where `exc.' and `inc.' refer to the exclusion and inclusion of `prompt data', meaning data within steep decay segments and flare intervals.

Regardless of whether or not data in flare intervals and steep decay segments are included, the same trend is observed over each successive dex bin requirement: the strength of the correlation increases, the degree of scatter reduces, and the sample size decreases. Furthermore, the linear regression slopes remain consistent with the full sample of regular LGRBs (N=382). These factors imply that using one method over the other for selecting well-sampled light curves is unlikely to strongly affect the main conclusions of this paper. However, for a given dex bin requirement, the sample size is smaller when excluding data within flare intervals and steep decay segments. For example, at $\geq5$ dex bins; N=102 when including these data compared to N=46 when excluding these data. This is because there are less dex bins to count over when these data (typically at early times) are excluded.

We opted to include the data within flare intervals and steep decay segments when selecting well-sampled light curves in Section \ref{sec:well-sampled} in order to obtain a larger sample size. Similarly, due to the trade off between the sampling requirements (number of dex bins) and the corresponding sample sizes, we chose the criteria of $\geq5$ dex bins to optimize the sampling requirements whilst maintaining a large sample size (N=102). We emphasize that using one method over the other is unlikely to strongly affect the main conclusions of this paper, provided a suitable number of dex bins are chosen.

\section{Distributions of redshift and observed flux for well-sampled regular LGRBs}
\label{sec:appendix well sampled lgrb distributions}
\begin{center}
\begin{figure}
\includegraphics[width=\columnwidth]{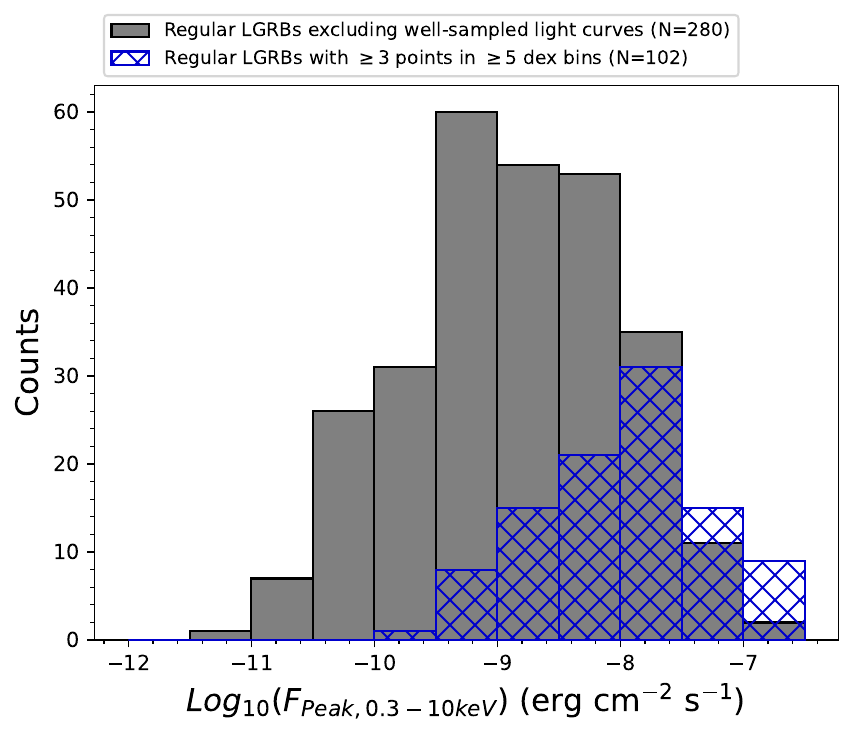}
    \caption{Distributions of the peak flux for each light curve in the \swift/XRT native band for our subsample of regular LGRBs divided into two groups: with and without well-sampled light curves. We calculate a KS statistic of 0.43 and a p-value of $\ll10^{-6}$, suggesting that the well-sampled regular LGRBs are biased towards higher fluxes compared to regular LGRBs which are not well-sampled.}
   \label{fig:flux_dist}
\end{figure}
\end{center}

We divide regular LGRBs into two groups: with and without well-sampled light curves. For each group, we examine the distributions of the peak (observed frame) X-ray flux in the \swift/XRT native band (including data in flare intervals and steep decay segments), as shown in Figure \ref{fig:flux_dist}. Well-sampled regular LGRBs are visibly offset towards higher peak fluxes than non-well-sampled regular LGRBs, and we calculate a KS statistic of 0.43 and a p-value of $\ll10^{-6}$, thus verifying the bias is strong, and statistically significant. The brighter peak fluxes in the observed frame are likely a cause of the increased sampling.

\begin{center}
\begin{figure}
\includegraphics[width=\columnwidth]{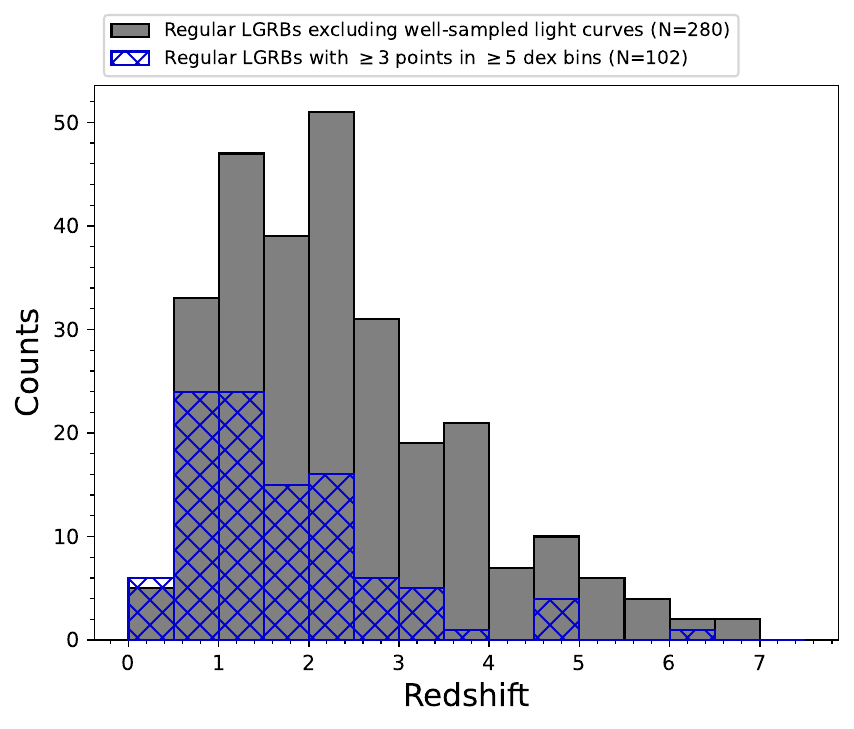}
    \caption{Distributions of redshift for our subsample of regular LGRBs divided into two groups: with and without well-sampled light curves. Regular LGRBs with well-sampled light curves are biased towards lower redshift compared to regular LGRBs without well-sampled light curves. However, the KS statistic of 0.26 indicates the difference is only moderate. The bias towards low redshift is likely related to better follow-up capabilities: a GRB of a given intrinsic brightness will have a brighter observed flux at lower redshift, and can thus be observed for longer, and with better sampling.}
   \label{fig:redshift_dist}
\end{figure}
\end{center}

Despite well-sampled regular LGRBs being biased towards GRBs with brighter X-ray fluxes in the observed frame, they are not biased in their rest frame X-ray brightness (see Section \ref{sec:well_samp_biases_main}). This indicates that well-sampled regular LGRBs are likely biased towards lower redshifts compared to regular LGRBs which are not well-sampled. We compare the redshift distributions of each group, as shown in Figure \ref{fig:redshift_dist}, and test for a bias to verify this. We calculate a KS statistic 0.26 and a p-value of $8.33\times10^{-5}$ between these two groups which indicate that, at a confidence interval of $\geq3\sigma$, our subsample of well-sampled regular LGRBs are biased towards lower redshifts compared to regular LGRBs without well-sampled light curves, as expected. The KS statistic indicates that the difference is moderate.

\section{The effect of light curve morphology on the correlation}
\label{sec:appendix morphology}
In Section \ref{sec:well_samp_biases_main}, we separated regular LGRBs into two groups: with and without canonical light curves, and established each group have consistent rest frame properties (X-ray brightness and isotropic gamma-ray energy\footnote{There is a small and subtle bias towards higher energies in canonical light curves}). This indicates that the improved correlation results (increased strength and reduced scatter) observed in well-sampled regular LGRBs are driven by increased sampling rather than a bias in the rest frame brightness and energy. However, in Section \ref{sec:morphology} we highlighted that the majority of well-sampled regular LGRBs, $\simeq70$ per cent, have canonical light curves. As a result, there is a possibility that it is the morphological types that drive the improved correlation results rather than the sampling. In this section, we investigate this possibility.

\begin{center}
\begin{figure*}
\includegraphics[width=1.6\columnwidth]{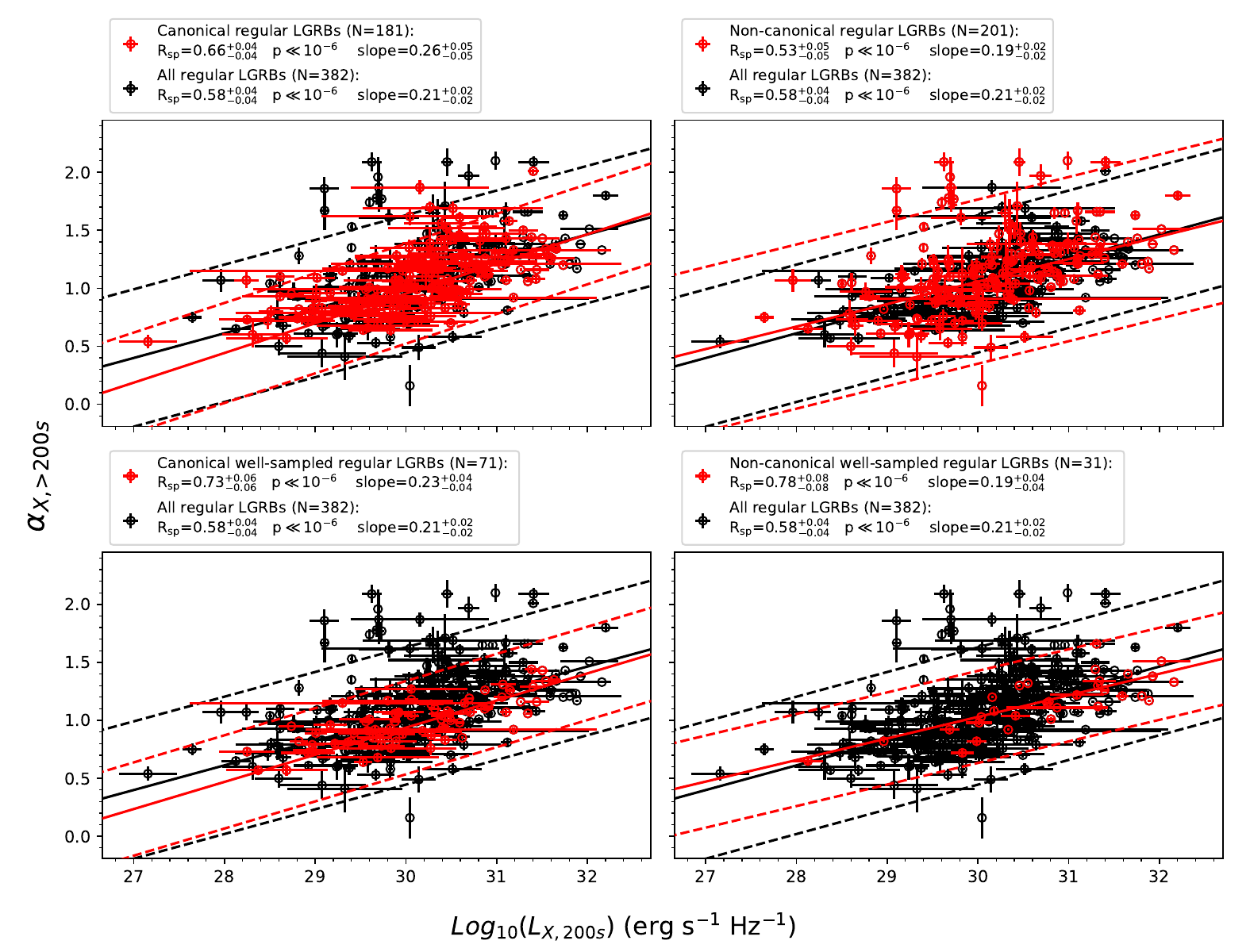}
    \caption{Plots of the correlation in different morphological types of regular LGRBs (left column: canonical, right column: non-canonical) before (top row) and after (bottom row) selecting well-sampled light curves. Regardless of morphological type, the correlation results become stronger and have less scatter after selecting light curves which are well-sampled. This supports the improved results being due to the sampling instead of the morphological type.}
   \label{fig:well_sampled_by_morph}
\end{figure*}
\end{center}

We separate regular LGRBs with well-sampled light curves into two groups: with and without canonical light curves, and test for the correlation separately in each. The correlation in these groups are shown in the bottom row of Figure \ref{fig:well_sampled_by_morph}. The results are consistent, at $\leq1\sigma$, between both groups with $R_\mathrm{sp}=0.73\pm 0.06$ and $R_\mathrm{sp}=0.78\pm0.08$ for canonical and non-canonical well-sampled regular LGRBs respectively, and are significantly stronger than observed in regular LGRBs. The slopes in each class are also consistent, at $\leq1\sigma$, with values of $0.23\pm0.04$ and $0.19\pm0.04$ for the canonical and non-canonical classes respectively. Furthermore, both classes contain the same (reduced) degree of scatter with a $2\sigma$ RMS value of 0.40. The improved correlation results in well-sampled but non-canonical light curves suggests the increased sampling drives the improved correlation results rather than the morphology.

We make the same split, this time to the parent regular LGRB sample (N=382), and test for the correlation in each as shown in the top row of Figure \ref{fig:well_sampled_by_morph}. We find that the correlation is stronger in the canonical group compared to the non-canonical group with $R_\mathrm{sp}=0.66\pm0.04$ and $R_\mathrm{sp}=0.53\pm0.05$ respectively, and their slopes are consistent within $1.4\sigma$ with values of $0.26\pm0.05$ and $0.19\pm0.02$ respectively. Furthermore, there is a larger degree of scatter in the non-canonical group with a $2\sigma$ RMS value of 0.71 compared to the a value of 0.43 in the canonical group. The improvement in correlation results in the canonical group is likely due to this group containing a larger fraction of well-sampled light curves, as the increase in correlation strength is not as significant as in either of the well-sampled groups (bottom row).

We therefore note that the improved correlation results appear to be directly related to the sampling rather than the morphology.

\section{The cause of the correlation changing at high redshift}
\label{sec:redshift_groups_luminosity_cut}
In Section \ref{sec:redshift_evolution_full}, we tested for the correlation in regular LGRBs separated into different redshift groups. The strength and the slope of the correlation in each redshift group are consistent with those observed in the full sample of regular LGRBs (N=382) at $\leq2\sigma$. However, in the highest redshift group, the fractional (and absolute) uncertainties on these parameters are the largest and the strength is the weakest. There are two possible means of causing this: the Malmquist bias meaning high redshift GRBs are biased towards a higher and narrower distribution of luminosities, or due to the correlation physically evolving with redshift. In this section, we investigate these scenarios.

\begin{center}
\begin{figure}
\includegraphics[width=\columnwidth]{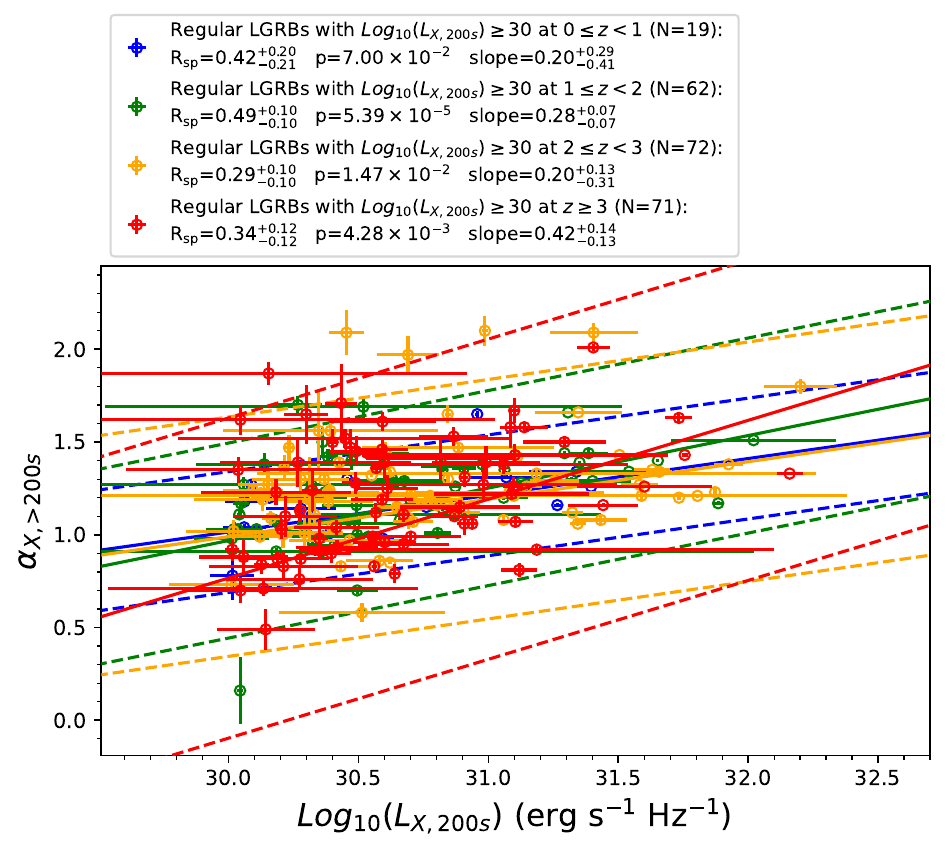}
    \caption{A comparison between the correlation in regular LGRBs separated into different redshift groups after selecting those with $\log_{10}$(\lum)$\geq 30$ erg s$^{-1}$ Hz$^{-1}$. The decreased robustness of the correlation is observed in each redshift group after the luminosity selection. This provides evidence that the decreased robustness of the correlation (as observed in high redshift GRBs in Section \ref{sec:redshift_evolution_full}) can be explained by a narrow range of higher luminosities. However, it does not rule out the possibility that the correlation may evolve with redshift.}
   \label{fig:redshift_evolution_figure_appendix}
\end{figure}
\end{center}

We reproduce the luminosity bias, observed in the high redshift group, in each redshift group by selecting GRBs with $\log_{10}$(\lum)$\gtrsim 30$ erg s$^{-1}$ Hz$^{-1}$. We then examine each redshift group for the correlation after reproducing the luminosity bias as shown in Figure \ref{fig:redshift_evolution_figure_appendix}. We compare the results in each redshift group before and after selecting high luminosity events. In the lowest redshift group, the strength of the correlation decreases and becomes largely uncertain, changing from $0.58\pm0.08$ to $0.42^{+0.20}_{-0.21}$, the significance decreases to $<3\sigma$, with a p-value change from $\ll10^{-6}$ to $7.00\times10^{-2}$. Furthermore, the slope becomes largely uncertain, changing from $0.16^{+0.03}_{-0.02}$ to $0.20^{+0.29}_{-0.41}$. A similar change in results is also observed in the other redshift groups. These collectively show that, regardless of redshift, a narrow range of high luminosities can explain the decreased robustness of the correlation that was observed in the highest redshift group in Section \ref{sec:redshift_evolution_full}.

Whilst this does not explicitly rule out that the correlation evolves with redshift, it provides evidence that the decreased robustness of the correlation at high redshift can be explained by the luminosity selection effect from the Malmquist bias. Observations of fainter GRBs at high redshift using more sensitive instruments would be required to potentially rule out the former scenario.

\section{The correlation in the full sample using later times}
\label{sec:later_time_full}
\begin{center}
\begin{figure}
\includegraphics[width=0.95\columnwidth]{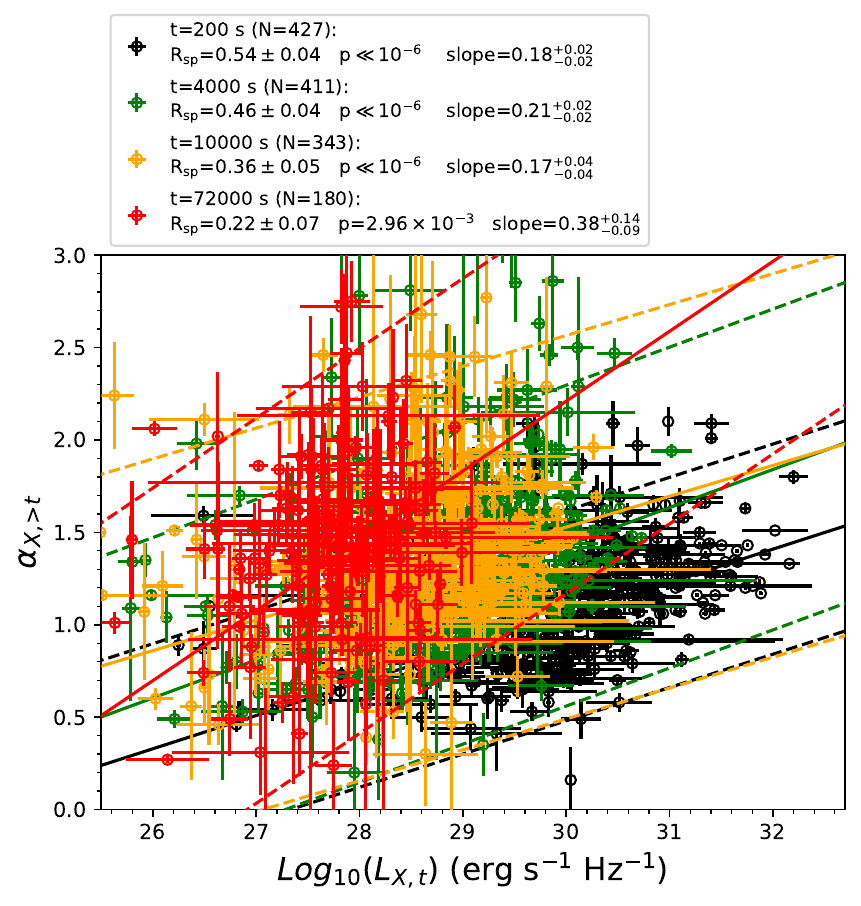}
    \caption{The same as in Figure \ref{fig:later_time_correlation}, with the luminosity measured at different times and the average rates of decay from each respective time, but this time for the full sample instead of for the subsample of well-sampled regular LGRBs. We re-analyse the light curves at each later time meaning that the sample size may change for each later time from excluding measurements with large errors, as in Section \ref{sec:measurements}.}
   \label{fig:later_time_full_fig}
\end{figure}
\end{center}

In Section \ref{sec:later times}, we examine the correlation in regular LGRBs with well-sampled light curves (N=102) as this subsample has the most robust results. Here, we repeat this but for the full sample (N=427) following the same method of re-analysing light curves at later times as in Section \ref{sec:later times}. Figure \ref{fig:later_time_full_fig} shows a plot of the luminosity and the average rate of decay measured using each later time for each light curve in our full sample (N=427). The correlation is observed with a statistical significance of $\geq3\sigma$ at all times. However, the strength of the correlation decreases with each later successive time and the slope increases along with the level of scatter.

\bsp	
\label{lastpage}
\end{document}